\documentclass[showpacs,preprintnumbers,amsmath,amssymb]{revtex4}
\usepackage{amsmath,amssymb,graphics,epsfig,subfigure}
\usepackage{color}

\begin{document}
\renewcommand{\baselinestretch}{1.3}

\title{Critical phenomena and chemical potential of charged AdS black hole}

\author{Shao-Wen Wei \footnote{weishw@lzu.edu.cn},
        Bin Liang \footnote{liangb2016@lzu.edu.cn}
        Yu-Xiao Liu \footnote{liuyx@lzu.edu.cn, corresponding author}}

\affiliation{Institute of Theoretical Physics, Lanzhou University, Lanzhou 730000, People's Republic of China}

\begin{abstract}
We study the thermodynamics and the chemical potential for a five-dimensional charged AdS black hole by treating the cosmological constant as the number of colors $N$ in the boundary gauge theory and its conjugate quantity as the associated chemical potential $\mu$. It is found that there exists a small-large black hole phase transition. The critical phenomena are investigated in the $N^{2}$-$\mu$ chart. In particular, in the reduced parameter space, all the thermodynamic quantities can be rescaled with the black hole charge such that these reduced quantities are charge-independent. Then we obtain the coexistence curve and the phase diagram. The latent heat is also numerically calculated. Moreover, the heat capacity and the thermodynamic scalar are studied. The result indicates that the information of the first-order black hole phase transition is encoded in the heat capacity and scalar. However, the phase transition point cannot be directly calculated with them. Nevertheless, the critical point linked to a second-order phase transition can be determined by either the heat capacity or the scalar. In addition, we calculate the critical exponents of the heat capacity and the scalar for the saturated small and large black holes near the critical point.
\end{abstract}



\maketitle

\section{Introduction}
\label{secIntroduction}

Thermodynamics of black holes continues to be an important and fascinating topic in gravitational physics. In particular, motivated by the AdS/CFT correspondence \cite{Maldacena,Gubser,Witten}, thermodynamics for anti-de Sitter (AdS) black holes has received great attention. Hawking and Page \cite{Hawking} found that, in an AdS space, there exists a phase transition between the stable large Schwarzschild black hole and the thermal gas. In the spirit of the correspondence, this phase transition is interpreted as the confinement/deconfinement phase transition in the dual strongly coupled gauge theory \cite{Witten2}. More interestingly, it was also found that there is a small-large black hole phase transition for charged or rotating AdS black holes \cite{Chamblin,Chamblin2,Caldarelli,Roychowdhury,Banerjee}.

Recently, the focus of the black hole thermodynamics is on the cosmological constant. In Refs.~\cite{Kastor,Dolan00,Cvetic}, the cosmological constant was treated as the thermodynamic pressure and its conjugate quantity as the thermodynamic volume of the AdS black hole. Then there is a precise pressure-volume oscillatory behavior, and the small-large black hole phase transition is identified with the liquid-gas phase transition of the van der Waals (vdW) fluid \cite{Kubiznak}. This has rejuvenated the study of the thermodynamics and critical phenomena for AdS black holes. Subsequently, the study was extended to other AdS black holes (for a recent review see \cite{Teo} and references therein).

On the other hand, from the AdS/CFT correspondence, the cosmological constant in the bulk corresponds to the number of colors in the gauge theory. Thus, it provides us another approach to understand the cosmological constant. In Ref. \cite{Dolan}, Dolan identified the cosmological constant with the number of colors and its conjugation as the chemical potential. He also calculated the chemical potential and found that, in the high temperature phase for the Yang-Mills theory, the chemical potential is negative and decreases with the black hole temperature. For the Schwarzschild-AdS black hole, its chemical potential approaches zero for a temperature lower than the Hawking-Page temperature, and changes its sign near the temperature of the divergent heat capacity. It was argued that such phenomenon may be used to understand the Bose-Einstein condensation in the dual field theory. The study was also generalized to other AdS black holes \cite{Maity}. Differently, the chemical potential is defined via the densities of the standard thermodynamic variables. Then the result showed that the chemical potentials for the five-dimensional Schwarzschild-AdS and charged-AdS black holes change sign precisely at the location of the Hawking-Page phase transition. While for the rotating AdS black hole with large angular frequency, the chemical potential change its sign in a stable black hole region with a temperature above the Hawking-Page temperature. Similar phenomenon was observed in the Gauss-Bonnet gravity \cite{Maity}. Furthermore, in Ref.~\cite{Traschen} the authors studied the associated chemical potential from the point of view of holographic entanglement entropy. The chemical potential and thermodynamic geometry were considered in Refs.~\cite{Zhang,Cai,Belhaj,Chabab}.

In the first law of the black hole thermodynamics, the number of colors and the chemical potential are a pair of the thermodynamic quantities. And varying them will trigger the black hole phase transition. Thus one of the aims of this paper is to study the thermodynamic critical phenomena in the number-chemical potential chart. The result shows that there exists a small-large black hole phase transition. Moreover we also discuss the relation between the phase transition and the heat capacity and the thermodynamic geometry. The study indicates that the information of the first-order and second-order black hole phase transitions is encoded in the heat capacity and thermodynamic scalar. However, only the second-order black hole phase transition point can be determined by them only.

The paper is organized as follows. In Sec. \ref{Classification}, we briefly review the thermodynamic quantities for the five-dimensional charged AdS black hole. The generalized Clapeyron equations are also obtained. In Sec. \ref{criticality}, the $N^{2}$-$\mu$ criticality is investigated. And the phase transition in the reduced parameter space is considered in Sec. \ref{reduce}. In Sec. \ref{heatgeometry}, the behaviors of the heat capacity and the thermodynamic geometry are examined. We also explore the question whether the information of the phase transition is encoded in the heat capacity or the thermodynamic geometry. The inverse question is also considered. After that, we numerically calculate the critical exponents for the heat capacity and thermodynamic scalar. Finally, the conclusions and discussions are given in Sec. \ref{Conclusion}.

\section{Thermodynamic quantities and first law}
\label{Classification}

For a five-dimensional charged AdS black hole, its line element can be expressed as
\begin{equation}
 ds_{5}^{2}=-f(r)dt^{2}+\frac{1}{f(r)}dr^{2}+r^{2}d\Omega_{3}^{2},
\end{equation}
where the metric function is \cite{Chamblin}
\begin{equation}
 f(r)=1-\frac{m}{r^{2}}+\frac{q^{2}}{r^{4}}+\frac{r^{2}}{l^{2}}.\label{metricfunction}
\end{equation}
This solution originates from the following action
\begin{equation}
 S=\frac{1}{16\pi G_{5}}\int_{M}d^{5}x\sqrt{-g}
   \left(R+\frac{12}{l^{2}}-l^{2}F^{2}\right).
\end{equation}
Such solution can also be uplift to ten dimensions \cite{Chamblin}
\begin{equation}
 ds^{2}_{10}=ds_{5}^{2}
      +l^{2}\sum_{i=1}^{3}\left(d\theta_{i}^{2}
        +\theta_{i}^{2}\left(d\varphi_{i}+\frac{2}{\sqrt{3}}A_{\nu}dx^{\nu}\right)^{2}\right),
        \label{ss}
\end{equation}
with $\nu$=0, 1, 2, 3, and 4. The variables $\theta_{i}$ are the direction cosines on $S^{5}$, which satisfy $\sum_{i=1}^{3}\theta_{i}^{2}$=1, and $\varphi_{i}$ are the rotation angles on $S^{5}$. The ten-dimensional spacetime (\ref{ss}) can be regarded as the near horizon geometry of the $N$ rotating black D3-branes in type IIB supergravity. Under such scenarios, the AdS radius $l$ is linked to the number $N$ of the D3-branes as \cite{Maldacena}
\begin{equation}
 l^{4}=\frac{\sqrt{2}Nl^{4}_{p}}{\pi^{2}}
\end{equation}
with $l_{p}$ the ten-dimensional Planck length. According to AdS/CFT correspondence, Eq. (\ref{ss}) can be viewed as the gravity dual to $\mathcal{N}=4$ supersymmetric Yang-Mills theory in the Coulomb branch. Thus the number $N$ is just the rank of the gauge group of the $SU(N)$ Yang-Mills theory. Since the number of degrees of freedom of the $\mathcal{N}=4$ supersymmetric Yang-Mills theory is in proportion to $N^{2}$ in the large $N$ limit \cite{Gubser2}, we will in the following investigation treat $N^{2}$ as a variable rather than $N$.

The parameters $m$ and $q$ appearing in the metric function (\ref{metricfunction}) are related to the black hole mass $M$ and charge $Q$ as
\begin{eqnarray}
 M=\frac{3\pi m}{8G_{5}},\quad
 Q=\frac{\sqrt{3}\pi q}{G_{5}l}.
\end{eqnarray}
By solving the equation $f(r)=0$, one can obtain the black hole horizon $r_{h}$. Then the black hole mass can be expressed as
\begin{eqnarray}
 M=\frac{3\pi}{8\pi G_{5}}\left(r_{h}^{2}+\frac{r_{h}^{4}}{l^{2}}
      +\frac{4G_{5}^{2}Q^{2}l^{2}}{3\pi^{2}r_{h}^{2}}\right).
\end{eqnarray}
According to the Bekenstein-Hawking entropy formula, the black hole entropy reads
\begin{eqnarray}
 S=\frac{\pi^{2}r_{h}^{3}}{2G_{5}}.
\end{eqnarray}
Here one needs to note the relation $G_{10}=\pi^{3}l^{5}G_{5}=l_{p}^{8}$. Using the `Euclidean trick', the black hole temperature is calculated as
\begin{eqnarray}
 T=\frac{1}{2\pi r_{h}}+\frac{r_{h}}{\pi l^{2}}
    -\frac{2G_{10}^{2}Q^{2}}{3l^{8}\pi^{9}r_{h}^{5}}.\label{t1}
\end{eqnarray}
The first law for the black hole is
\begin{equation}
 dM=TdS+\Phi dQ+\mu dN^{2}.
\end{equation}
And the thermodynamic quantities can be found
\begin{eqnarray}
 T&=&\left(\frac{\partial M}{\partial S}\right)_{Q, N^{2}}
   =\frac{-\sqrt{2}\pi NQ^2+3\pi^2(NS)^{4/3}+6S^2}
     {6\sqrt[8]{2}\pi^{7/4}N^{11/12}S^{5/3}},\label{t2}\\
 \Phi&=&\left(\frac{\partial M}{\partial Q}\right)_{S, N^{2}}
   =\frac{\sqrt[12]{N} Q}{2^{5/8}\pi^{3/4}S^{2/3}},\\
 \mu&=&\left(\frac{\partial M}{\partial N^{2}}\right)_{S, Q}
   =\frac{\sqrt{2}\pi NQ^2+15\pi^2(NS)^{4/3}-33S^2}
     {96\sqrt[8]{2}\pi^{7/4}N^{35/12}S^{2/3}},\label{mu}
\end{eqnarray}
where we have set $l_{p}=1$. Some notes should be clearly presented here for the chemical potential $\mu$. From the point of view of the AdS/CFT correspondence, chemical potential is a focused problem. Recently, Dolan \cite{Dolan} showed a chemical potential, which thermodynamically conjugates to the color number, for the neutral black hole in $AdS_{5}$. He also found that the chemical potential approaches zero as the temperature is lowered below the Hawking-Page temperature. The study of the chemical potential was also extended to other black hole in Ref. \cite{Maity}. These results, to some extent, indicate that the chemical potential behaves very different from that for a vdW fluid. Our suggestion is that although such chemical potential is defined as the conjugate quantity to the color number, it has different meanings as the one of the vdW fluid. So comparison between them should be very careful. Here we only treat it as a conjugate quantity, such like the electric potential. Of course, the study of that chemical potential will shine some lights on the boundary gauge theory living in an AdS space.

It is easy to check that the temperature $T$ obtained with the `Euclidean trick' (\ref{t1}) is the same as the one obtained with the first law (\ref{t2}). Correspondingly, we have the smarr formula
\begin{equation}
 7M=8TS+6Q\Phi+8\mu N^{2}.
\end{equation}
Next, we would like to consider the phase transition in different parameter spaces. In Ref. \cite{Cai}, the authors considered the thermodynamic properties of the density of these thermodynamic quantities. Here we expect to investigate the thermodynamics for the whole black holes, especially for the phase transition between the small and large black holes. So we will not consider the density of these thermodynamic quantities. Using these thermodynamic quantities, the Helmholtz free energy can be expressed in the form
\begin{equation}
 \mathcal{F}=M-TS
    =\frac{5\sqrt{2}\pi N Q^2+3\pi^2 (N S)^{4/3}-3S^2}
    {12\sqrt[8]{2}\pi^{7/4}N^{11/12}S^{2/3}}.
\end{equation}
Differentiating it, we get
\begin{equation}
 d\mathcal{F}=-SdT+\mu dN^{2}+\Phi dQ.\label{dfree}
\end{equation}
At the point of the first-order phase transition of A phase and B phase, we have
\begin{equation}
 T_{A}=T_{B},\quad N^{2}_{A}=N^{2}_{B}, \quad Q_{A}=Q_{B},
\end{equation}
or, equivalently, $\mathcal{F}_{A}=\mathcal{F}_{B}$. Similar to Ref. \cite{WeiLiu}, there are several cases to explore the phase transition.\\

\noindent\textbf{Case I:} $N^{2}$ and $Q$ fixed. Integrating Eq. (\ref{dfree}),
\begin{equation}
 \int_{T_{A}}^{T_{B}}S dT=0.
\end{equation}

\noindent\textbf{Case II:} $T$ and $Q$ fixed. It gives
\begin{equation}
 \int_{N^{2}_{A}}^{N^{2}_{B}}\mu dN^{2}=0.
\end{equation}

\noindent\textbf{Case III:} $N^{2}$ and $T$ fixed. One has
\begin{equation}
 \int_{Q_{A}}^{Q_{B}}\Phi dQ=0.
\end{equation}

According to these three cases, we can construct three equal area laws along $T$-$S$, $N^{2}$-$\mu$, and $Q$-$\Phi$ lines. The second-order phase transition can be, equivalently, determined by
\begin{eqnarray}
 (\partial_{S}T)_{N^{2},Q}&=&(\partial_{S,S}T)_{N^{2},Q}=0,\\
 (\partial_{\mu}N^{2})_{T,Q}&=&(\partial_{\mu,\mu}N^{2})_{T,Q}=0,\label{condition2}\\
 (\partial_{\Phi}Q)_{N^{2},T}&=&(\partial_{\Phi,\Phi}Q)_{N^{2},T}=0.
\end{eqnarray}
Correspondingly, along the coexistence curve, the generalized Clapeyron equations are
\begin{eqnarray}
 \bigg(\frac{dN^{2}}{dT}\bigg)_{Q}&=&\frac{S_{B}-S_{A}}{\mu_{B}-\mu_{A}},\label{Clapeyron}\\
 \bigg(\frac{dQ}{dN^{2}}\bigg)_{T}&=&\frac{\mu_{B}-\mu_{A}}{\Phi_{B}-\Phi_{A}},\label{Clapeyron2}\\
 \bigg(\frac{dT}{dQ}\bigg)_{N^{2}}&=&\frac{\Phi_{B}-\Phi_{A}}{S_{B}-S_{A}}.\label{Clapeyron3}
\end{eqnarray}
In addition, it is important to note that the oscillatory behaviors and Clapeyron equations discussed here require that $\mu$, $T$, and $\Phi$ are regular functions in the expected parameter range.

\section{$N^{2}$-$\mu$ criticality}
\label{criticality}

\begin{figure}
\center{\subfigure[]{\label{nmunmu1a}
\includegraphics[width=6cm]{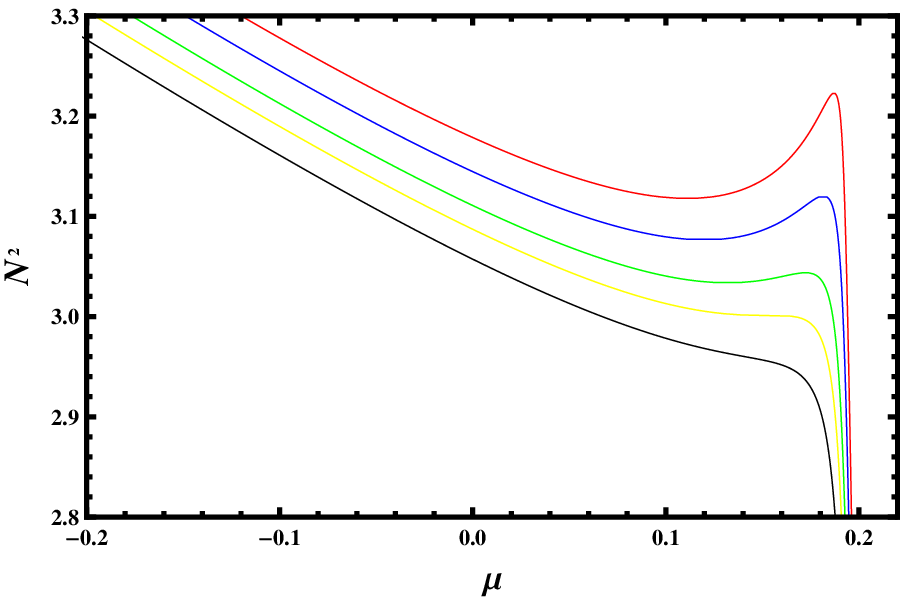}}
\subfigure[]{\label{nmunmu1b}
\includegraphics[width=6cm]{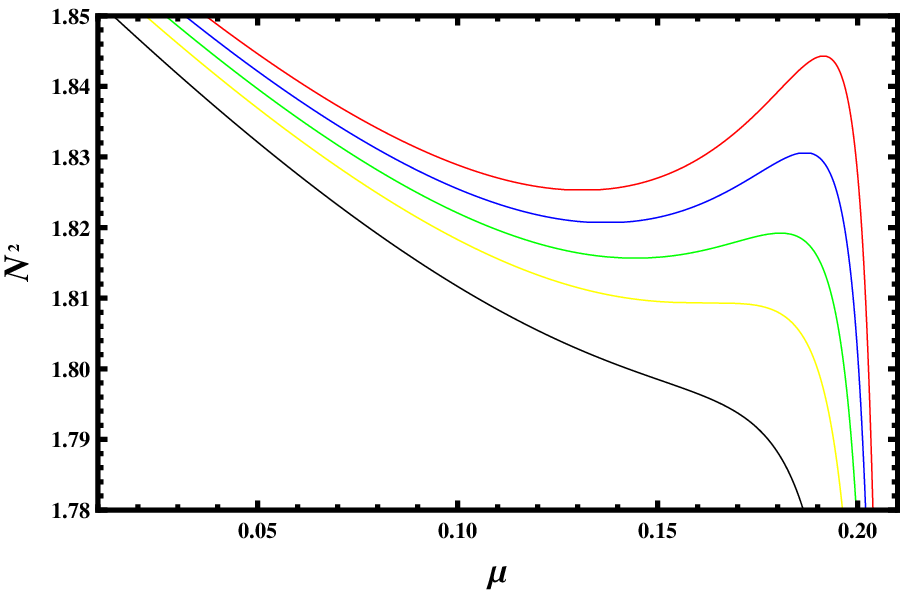}}}
\caption{$N^{2}$-$\mu$ behavior. (a) $T$=0.4680, 0.4685, 0.4690, 0.4694(=$T_{c}$), and 0.4698 from top to bottom with fixed charge $Q$=5. (b) $Q$=3.39, 3.40, 3.41, 3.4210(=$Q_{c}$), and 3.44 from top to bottom with fixed temperature $T$=0.5.}\label{nmunmu}
\end{figure}

\begin{figure}
\center{\subfigure[]{\label{nmuextremalA}
\includegraphics[width=6cm]{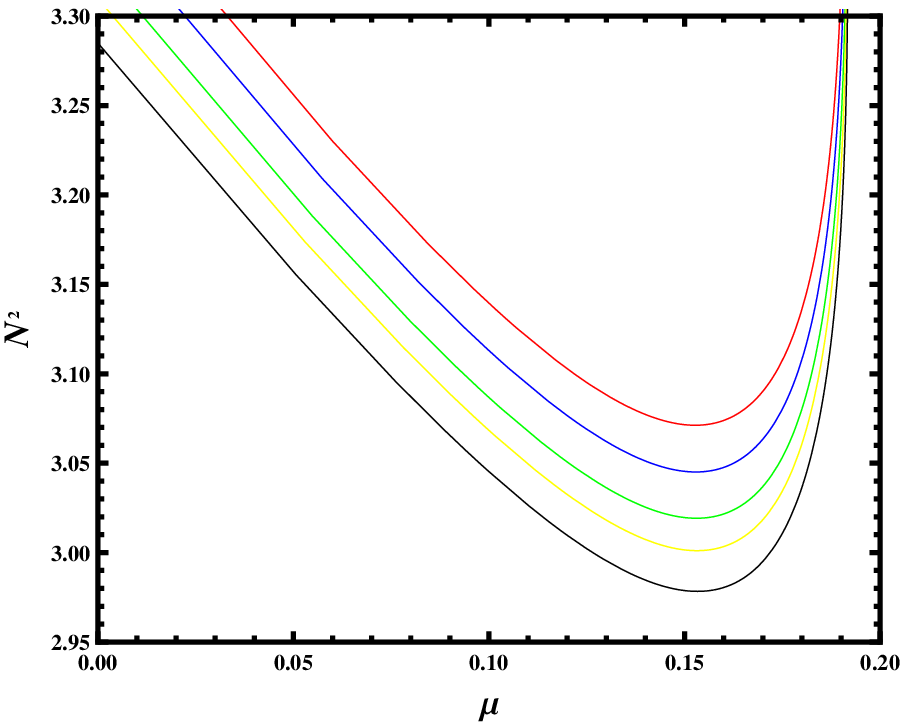}}
\subfigure[]{\label{nmuextremalB}
\includegraphics[width=6cm]{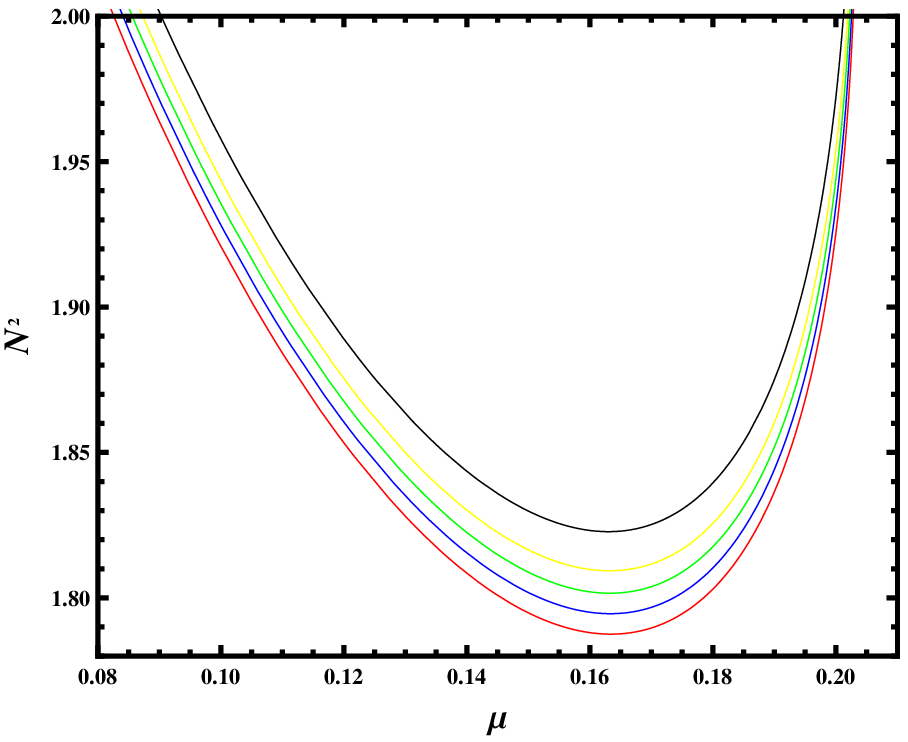}}}
\caption{Extremal point in the $N^{2}$-$\mu$ chart. (a) $T$=0.4680, 0.4685, 0.4690, $T_{c}$=0.4694, and 0.4698 from top to bottom. (b) $Q$=3.39, 3.40, 3.41, $Q_{c}$=3.4210, and 3.44 from bottom to top.}\label{nmuextremalAB}
\end{figure}

Here we are interested in the $N^{2}$-$\mu$ criticality. With the help of Eq. (\ref{mu}), we present the behavior of the number of colors $N^{2}$ as a function of the chemical potential $\mu$ in Fig.~\ref{nmunmu}. Figure \ref{nmunmu1a} shows the $N^{2}$-$\mu$ behavior when the charge $Q$=5 while the temperature increases from 0.4680 to 0.4698. For low temperature, there clearly exhibits an oscillatory behavior, which implies a first-order phase transition. While for high temperature, such oscillatory behavior disappears and no phase transition occurs. For fixed temperature $T$, the behavior is plotted in Fig.~\ref{nmunmu1b}. We can see that the oscillatory behavior occurs for small charge, while disappears for large charge. According to this oscillatory behavior, the critical point can be determined by the conditions (\ref{condition2}) as follows:
\begin{eqnarray}
 N_{c}^{2}=\frac{3\times 5^{2/3}\times 6^{1/3}}{\pi^{10/3}}Q^{\frac{4}{3}},\quad
 \mu_{c}=\frac{7\pi^{8/3}}{480\times5^{1/12}\times6^{1/6}}Q^{-\frac{1}{6}},\quad
 T_{c}=\frac{2\times2^{5/6}(3/\pi)^{1/3}}{5\times5^{1/12}}Q^{-\frac{1}{6}}.
\end{eqnarray}
For each oscillatory curve, there are two extremal points, and they coincide with each other at the critical point. Here, we show the extremal point in the $N^{2}$-$\mu$ chart in Fig.~\ref{nmuextremalAB} for fixed temperature and charge, respectively. An interesting phenomenon is that the number of colors $N^{2}$ has a minimum, below which there is no oscillatory behavior, and no phase transition occurs. From the figure, one can grasp that the minimum of $N^{2}$ increases with the charge $Q$, while decreases with the temperature $T$.

\begin{figure}
\includegraphics[width=8cm]{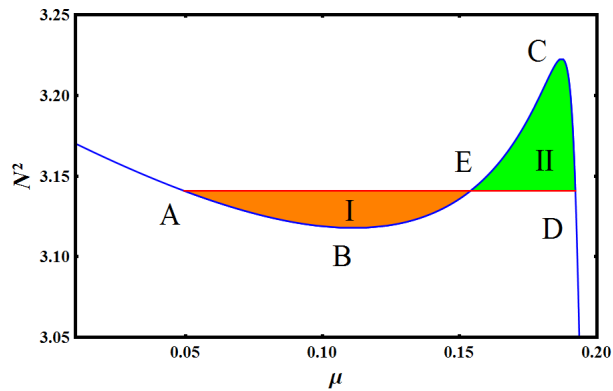}
\caption{Equal area law with $T$=0.468 and $Q$=5.}\label{pnmuequalarealaw}
\end{figure}

Along each oscillatory curve showed in Fig.~\ref{nmunmu}, these parts of positive slope are unphysical, and should be excluded for a thermodynamic process. Such mechanism is the Maxwell equal area law. Taking for an example, we construct the law on the isotherm with $T=0.468$ and $Q=5$, which is displayed in Fig.~\ref{pnmuequalarealaw}. The line BC with positive slope is excluded by constructing the two equal areas I and II. Then the oscillatory curve ABCD is replaced by a horizontal line AED. The horizontal line shares the same value of $N^{2}$=3.0011, which corresponds to the phase transition point. In fact, the lines AB and CD are also physical branches, and these black hole states will exist for some fine-tuning cases. Since the large black hole has the small chemical potential, we can call the line AB as the super low color number large black hole branch, and the line CD as the super high color number small black hole branch.

\begin{figure}
\center{\subfigure[]{\label{pFreeenergyN}
\includegraphics[width=6cm]{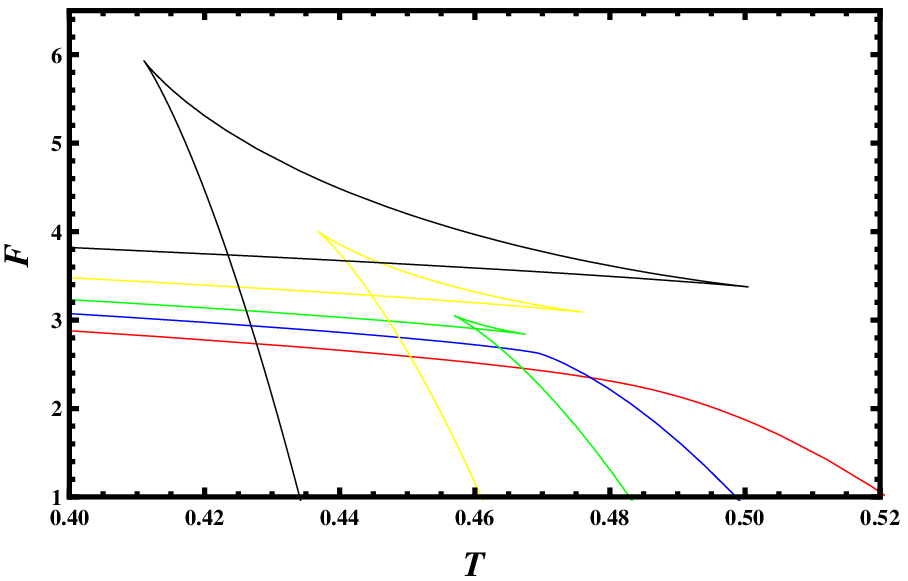}}
\subfigure[]{\label{pFreeEnergyQ}
\includegraphics[width=6cm]{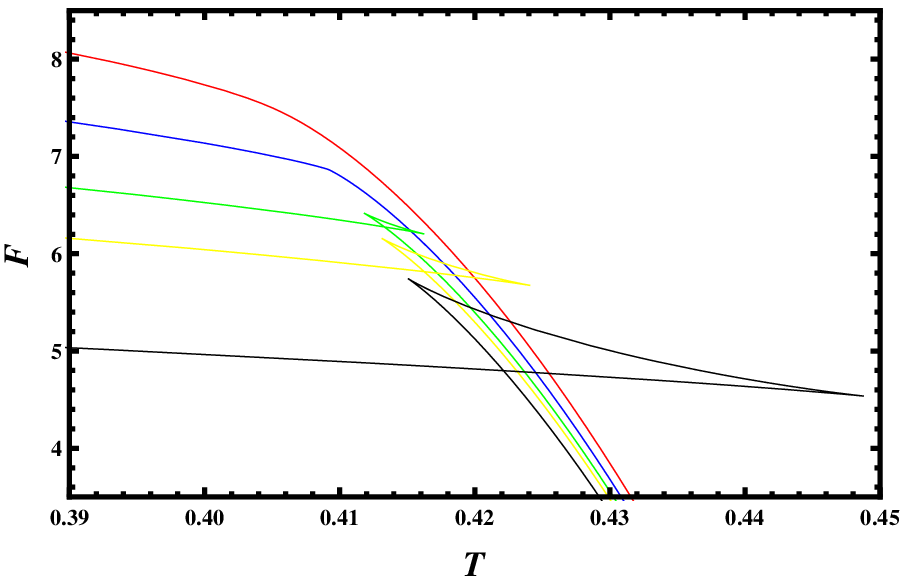}}}
\caption{Helmholtz free energy $\mathcal{F}$. (a) $N^{2}$=2, 3.0011(=$N^{2}_{c}$), 4, 6, and 10 from bottom to top with the charge $Q$=5. (b) $Q$=13, 11.3944(=$Q_{c}$), 10, 9, and 7 from top to bottom with fixed number of color $N^{2}$=9.}\label{pFreeenergy}
\end{figure}

On the other hand, the phase transition point can also be determined by the Helmholtz free energy. The behavior of the Helmholtz free energy is described in Fig.~\ref{pFreeenergy}. When fixed charge $Q$=5, the free energy is plotted in Fig.~\ref{pFreeenergyN} for $N^{2}$=2, 3.0011, 4, 6, and 10 from bottom to top. For large $N^{2}\geq 3.0011$, there exhibits a swallow tail behavior. Since a system prefers a state of low free energy, such behavior indicates a phase transition of the first-order occurs at the intersection point, where the free energy does not change smoothly. When decreasing $N^{2}$, the swallow tail behavior shrinks, and tends to vanish at $N^{2}$=3.0011, where the first-order phase transition becomes a second-order one. Further decreasing $N^{2}$, the swallow tail behavior of the free energy completely disappears. And it will smoothly decrease with the temperature. In Fig.~\ref{pFreeEnergyQ}, we keep the number of colors $N^{2}$=9, while vary the charges $Q$ from 13 to 7. The result indicates that the phase transition is inclined to occur for small charge. With a simple calculation, one can obtain the critical point of the free energy
\begin{eqnarray}
 \mathcal{F}_{c}=\frac{3^{1/6}\times5^{7/12}}{2\times2^{5/6}\pi^{2/3}}Q^{\frac{7}{6}}.
\end{eqnarray}

\begin{figure}
\center{\subfigure[]{\label{pnmupt}
\includegraphics[width=6cm]{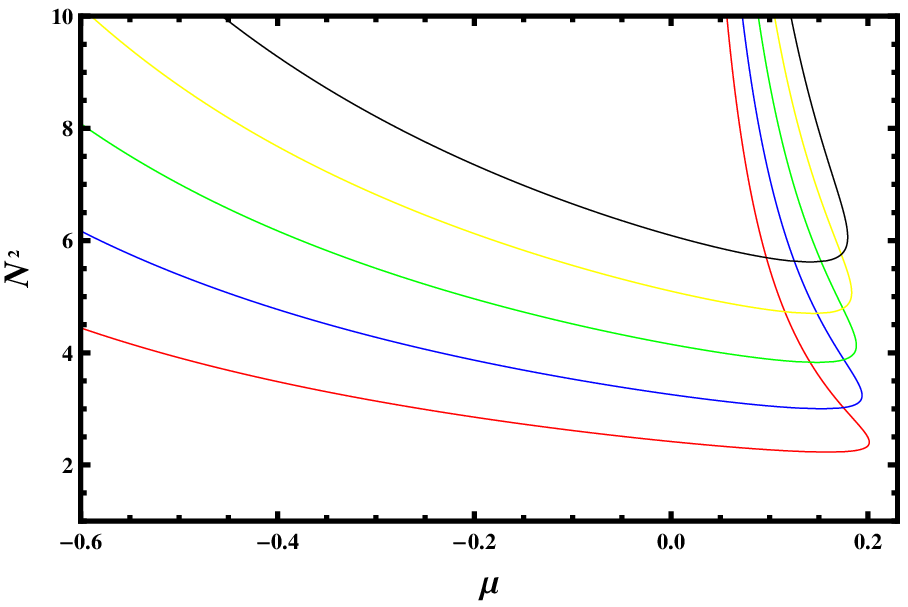}}
\subfigure[]{\label{pnmunt}
\includegraphics[width=6cm]{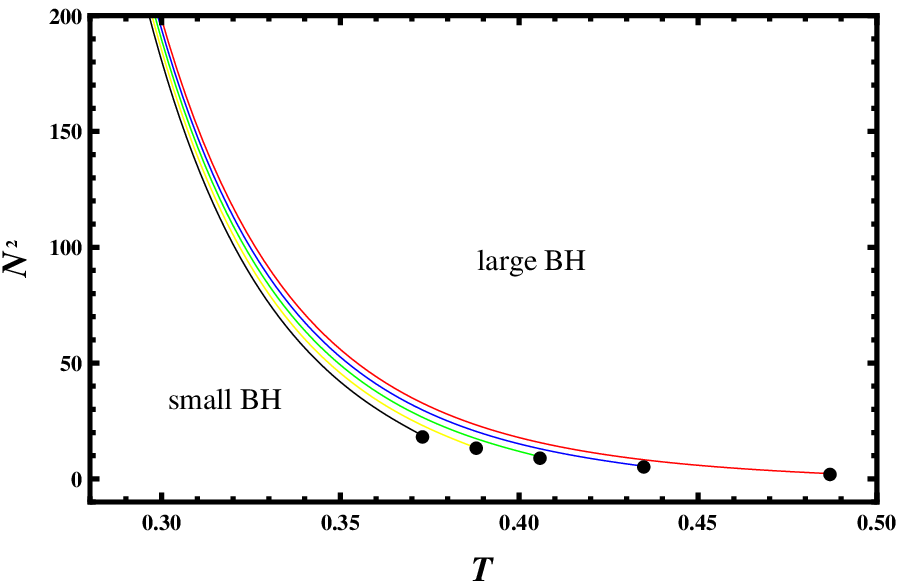}}}
\caption{(a) $N^{2}$-$\mu$ phase diagram. The charge is set to $Q$=4, 5, 6, 7, and 8 from bottom to top. Above the curve is the coexistence region of small and large black holes. Left is the large black hole phase and right is the small black hole phase. (b) $N^{2}$-$T$ phase diagram. The charge is set to $Q$=4, 8, 12, 16, and 20 from right to left.}\label{pnmuptaa}
\end{figure}

Then varying the parameter $T$ or $Q$, we can obtain the $N^{2}$-$\mu$ phase diagram through constructing the Maxwell equal area law or determining the intersection point of the swallow tail behavior of the free energy. The coexistence curve is presented in Fig.~\ref{pnmupt} for the charge $Q$=4, 5, 6, 7, and 8. The coexistence region of small and large black holes locates above the curve. And the region of large black hole phase locates at the left side of the curve, while the small black hole phase is at the right side. The region below the critical point is the super-critical region, in which the large and small black holes cannot be clearly distinguished. Moreover, we can also plot the phase diagram in the $N^{2}$-$T$ chart, see Fig.~\ref{pnmunt}. The black dots denote the critical points for different charge $Q$. It is clear that low phase transition temperature requires a large color number $N^{2}$. One can also see that, for different charge $Q$, the coexistence curve only has a tiny difference.

\begin{figure}
\includegraphics[width=8cm]{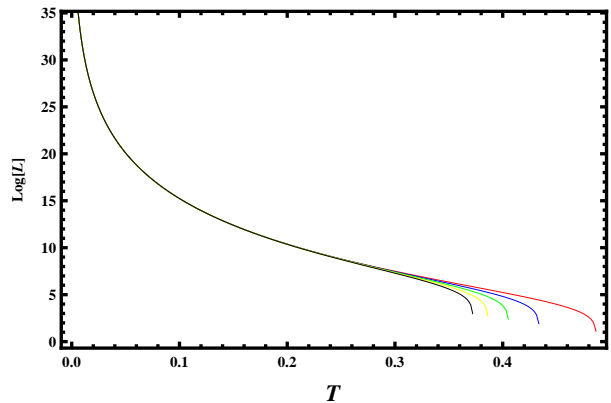}
\caption{Latent heat as a function of the temperature $T$. The charge $Q$=4, 8, 12, 16, and 20 from right to left. Since near the critical point, one has $\log{L}$=-$\infty$, so we adopt a cut off for the latent heat.}\label{platentheat}
\end{figure}

As is well known, the black hole system will suffer a latent heat absorbed or released during the first-order phase transition. While the latent heat vanishes when the system passes the second-order phase transition point \cite{WeiLiu2}. This provides us a possible method to distinguish the first and second-order phase transitions. Here, it is clear that the critical point is a second-order phase transition point. While the coexistence curve is a first-order one, and thus the latent heat may be remarkable when the system crosses it. Combining with the Clapeyron equation (\ref{Clapeyron}), the latent heat paid by the system to cross the coexistence curve can be calculated with the following formula:
\begin{eqnarray}
 L=T\Delta\mu \left(\frac{dN^{2}}{dT}\right)_{Q},
\end{eqnarray}
where $\Delta\mu$ is the difference of the chemical potentials between the small and large black hole phases. With the numerical calculation, we obtain the latent heat, which is plotted in Fig.~\ref{platentheat}. For different charge, the latent heat $L$ shares the similar behavior. First, it starts from a very large value for a small temperature $T$, then rapidly decreases with the temperature, and finally decreases to zero at the critical temperature. Such image again confirms that the critical point of the phase transition is a second-order one.

\section{Reduced parameter space}
\label{reduce}

\begin{figure}
\center{\subfigure[]{\label{pReducedNmu}
\includegraphics[width=6cm]{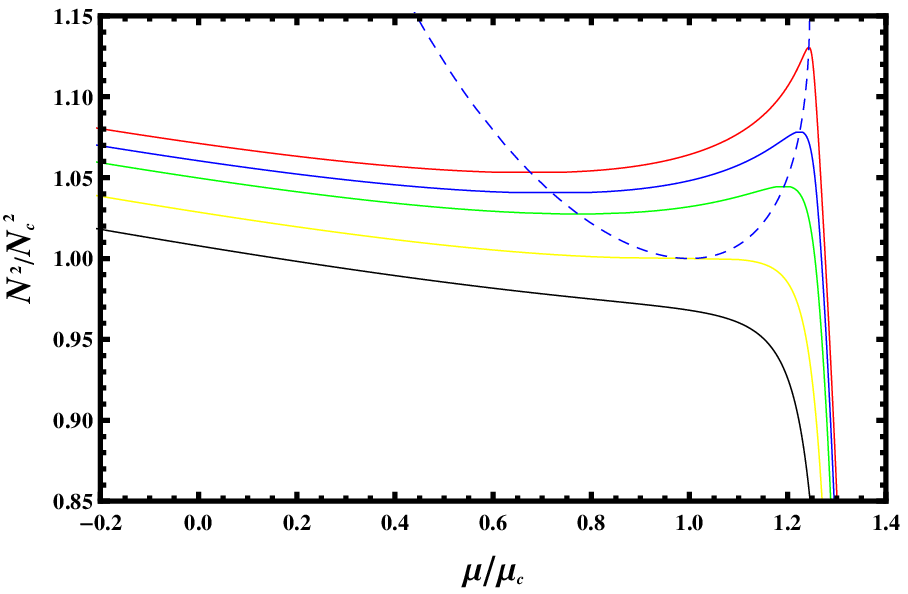}}
\subfigure[]{\label{pReducedF}
\includegraphics[width=6cm,height=4cm]{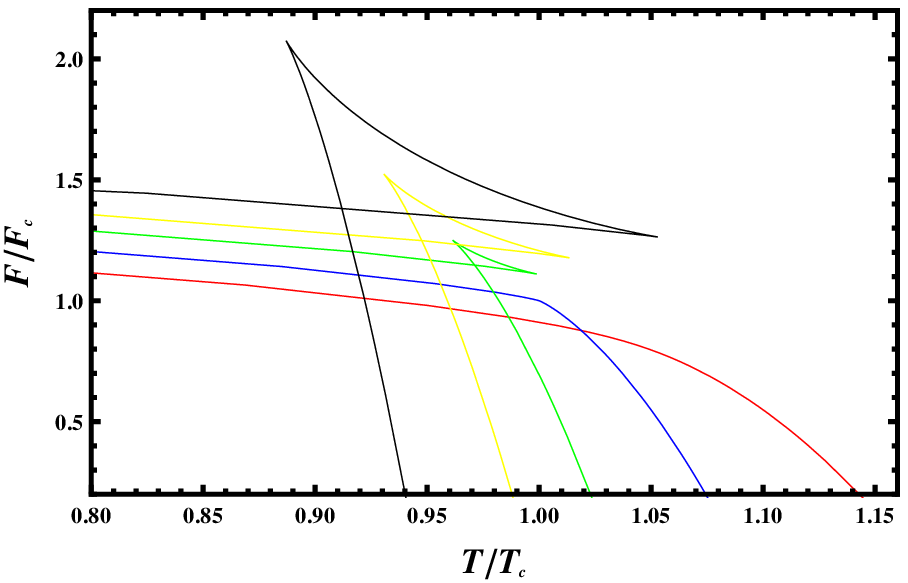}}}
\caption{(a) $\tilde{N}^{2}$-$\tilde{\mu}$ behavior for $\tilde{T}=T/T_{c}$=0.996, 0.997, 0.998, 1, and 1.002 from top to bottom. Blue dashed line denotes the extremal point line. (b) Free energy $\mathcal{F}$ for $\tilde{N}^{2}$=0.6, 1, 1.5, 2, and 3 from bottom to top.}\label{preduced}
\end{figure}

\begin{figure}
\center{\subfigure[]{\label{pReducedNTphase}
\includegraphics[width=6cm]{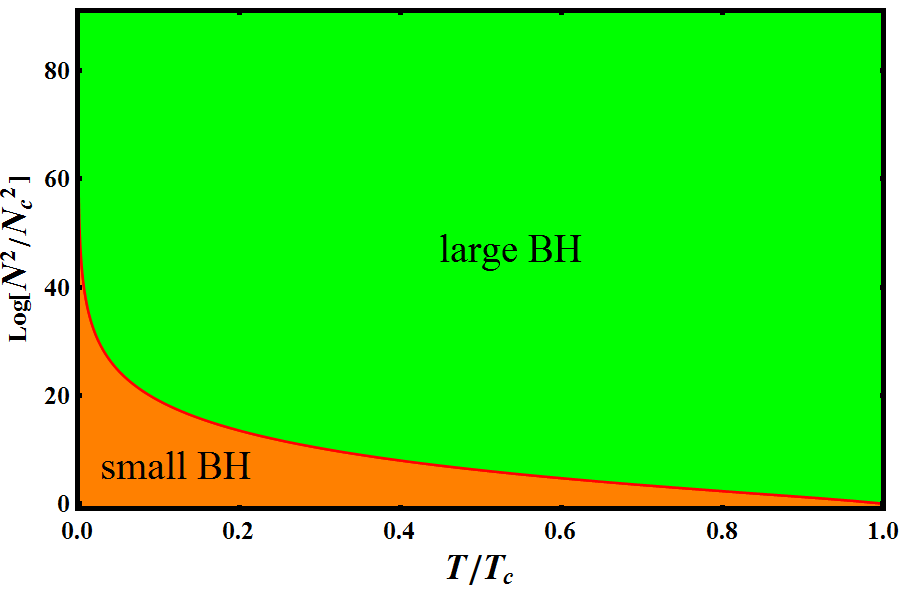}}
\subfigure[]{\label{pReducedNmuphase}
\includegraphics[width=6cm]{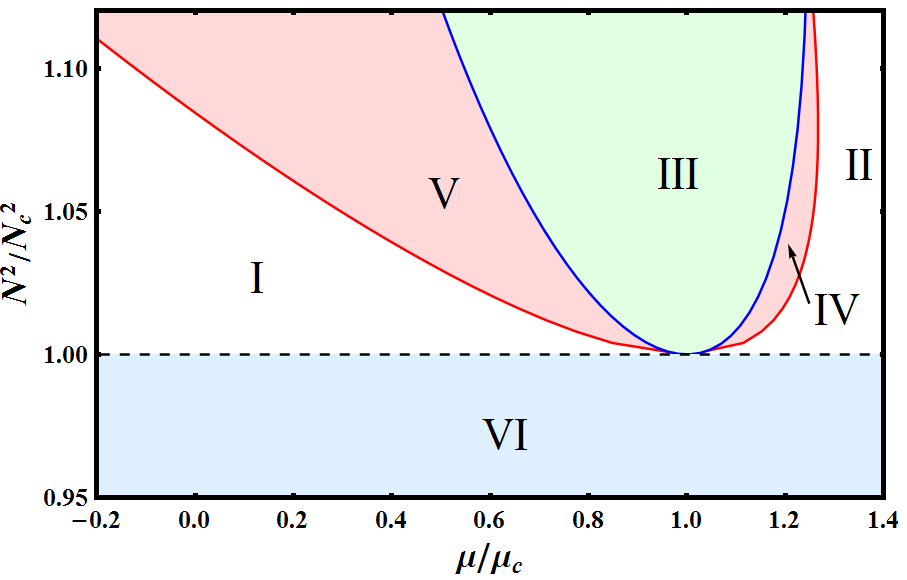}}
}
\caption{Phase structures in the reduced parameter space. (a) $\tilde{N}^{2}$-$\tilde{T}$ phase diagram. (b) $\tilde{N}^{2}$-$\tilde{\mu}$ phase diagram.} \label{ppReducedNTphase}
\end{figure}

\begin{figure}
\includegraphics[width=8cm]{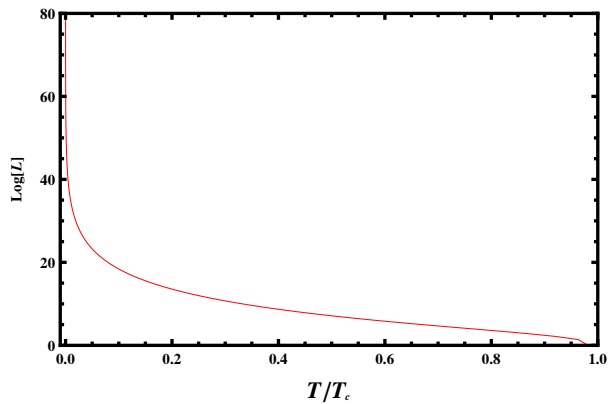}
\caption{Latent heat as a function of the temperature $\tilde{T}$ in the reduced parameter space.}\label{pReducedHeat}
\end{figure}

From above, one can see that the critical point and phase diagram closely depend on the black hole charge $Q$. Similar to the vdW fluid, it is interesting to examine the black hole thermodynamics in the reduced parameter space, where we can make these thermodynamic quantities dimensionless. Therefore, reduced quantities can be defined as $\tilde{A}=A/A_{c}$ with $A_{c}$ being the critical value. After doing this, we list these reduced thermodynamic quantities,
\begin{eqnarray}
 \tilde{T}&=&\frac{10\tilde{S}^{2}+15\tilde{N}^{4/3}\tilde{S}^{4/3}-\tilde{N}}{24\tilde{N}^{11/12}\tilde{S}^{2/3}},\\
 \tilde{\mathcal{F}}&=&\frac{-\tilde{S}^{2}+3\tilde{N}^{4/3}\tilde{S}^{4/3}+\tilde{N}}{3\tilde{N}^{11/12}\tilde{S}^{2/3}},\\
 \tilde{\mu}&=&\frac{-55\tilde{S}^{2}+75\tilde{N}^{4/3}\tilde{S}^{4/3}+\tilde{N}}{21\tilde{N}^{35/12}\tilde{S}^{2/3}}.
\end{eqnarray}
Obviously, these reduced quantities are charge-independent. So in the reduced parameter space, we decrease one free parameter.

In Fig.~\ref{pReducedNmu}, we give the behavior of the reduced color number $\tilde{N}^{2}$ against the reduced chemical potential $\tilde{\mu}$. Clearly, there will be the oscillatory behavior for the reduced temperature $\tilde{T}<1$, and that behavior disappears for $\tilde{T}>1$. Also, with the decrease of $\tilde{T}$, the oscillatory behavior becomes notable, and the reduced color number takes large value. Correspondly, it reveals a swallow tail behavior of the reduced free energy for $\tilde{T}<1$, see Fig.~\ref{pReducedF}.

With the help of the swallow tail behavior, we get the phase diagram showed in Fig.~\ref{ppReducedNTphase}. In Fig.~\ref{pReducedNTphase}, we show the phase diagram in the reduced $\tilde{N}^{2}$-$\tilde{T}$ chart. The red solid line is the coexistence line of the small and large back holes. The regions below and above the coexistence line are the small and large black hole phases, respectively. We can also find that, the phase transition value of $\tilde{N}^{2}$ grows exponentially when $\tilde{T}$ approaches zero. Moreover, we display the $\tilde{N}^{2}$-$\tilde{\mu}$ phase diagram in Fig.~\ref{pReducedNmuphase}. It is clear that six black hole phases emerge. The red solid line is the phase transition curve line. Regions I and II denote the large and small black hole phases, respectively. Region III is the coexistence region of small and large black holes. Regions IV and V are two metastable regions. After considering their nature, we can name them the super high color number small black hole phase and super low color number large black hole phase, respectively. The last region VI is the super-critical black hole region. One of its property is that the small and large black holes cannot be clearly distinguished, and thus no phase transition occurs. Analogy to the vdW fluid, we can name the boundary of regions I and V as the saturated large black hole curve and that of regions II and IV as the saturated small black hole curve.

The latent heat is depicted in Fig.~\ref{pReducedHeat}. It is similar to the one showed in Fig.~\ref{platentheat}. At the reduced critical temperature $\tilde{T}=1$, the latent heat vanishes, indicating a second-order phase transition. Note that at the critical temperature, $\log[L]=-\infty$, and we have adopted a cut off for it.

\section{Heat capacity and thermodynamic geometry}
\label{heatgeometry}

In this section, we will turn to another two black hole thermodynamic quantities, the heat capacity and thermodynamic scalar, which are believed to have important implications in revealing the black hole phase transition. Especially, the divergent points of them are generally thought to relate with the phase transition points.

\subsection{Heat capacity and phase transition}

\begin{figure}
\center{\subfigure[]{\label{pCapacitysa}
\includegraphics[width=6cm]{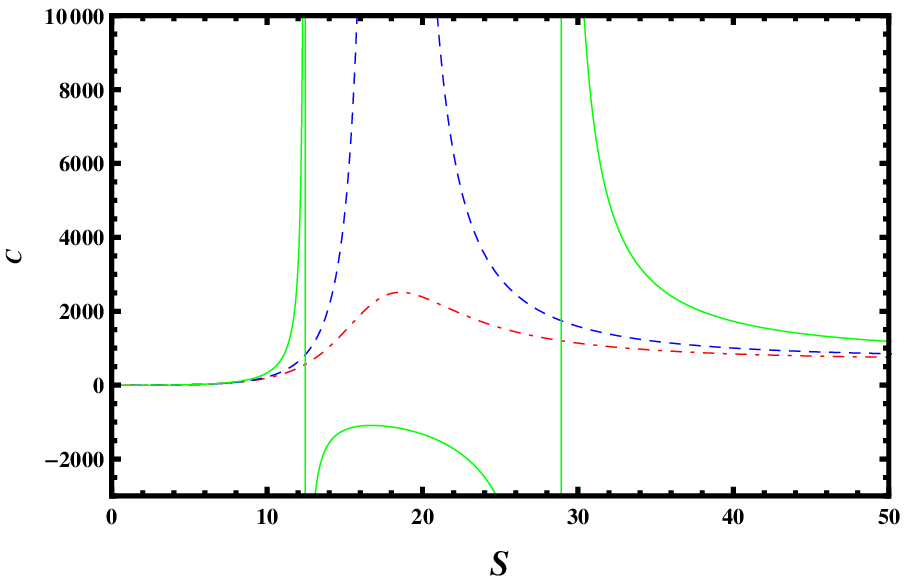}}
\subfigure[]{\label{pCapacityT}
\includegraphics[width=6cm]{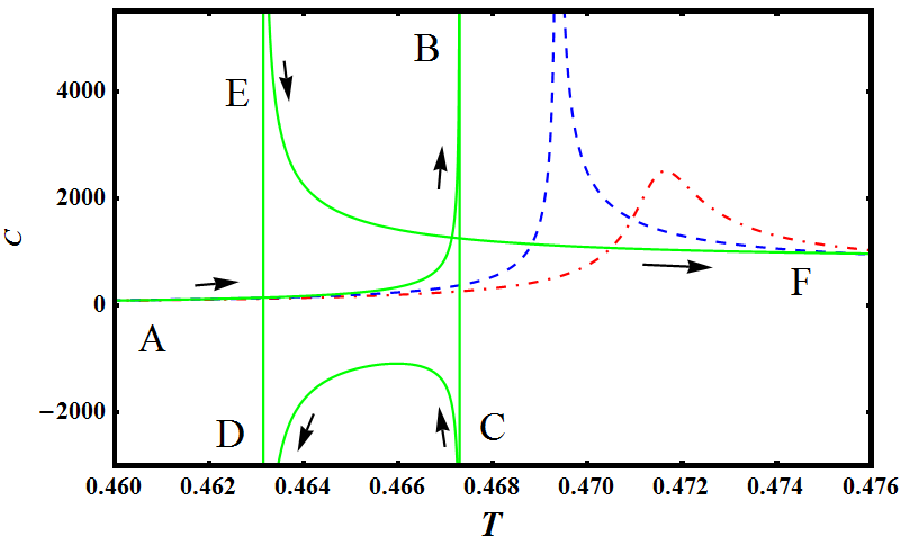}}}
\caption{Heat capacity with $Q$=5. (a) $C$ vs. $S$ and (b) $C$ vs. $T$. The color number is set to $N^{2}$=2.8, 3.0011, and 3.5 for red dot dashed line, blue dashed line, and green solid line respectively. The arrows indicate the increase of the entropy $S$.}\label{ppCapacity}
\end{figure}

\begin{figure}
\center{\subfigure[]{\label{pCapacityPHL}
\includegraphics[width=6cm]{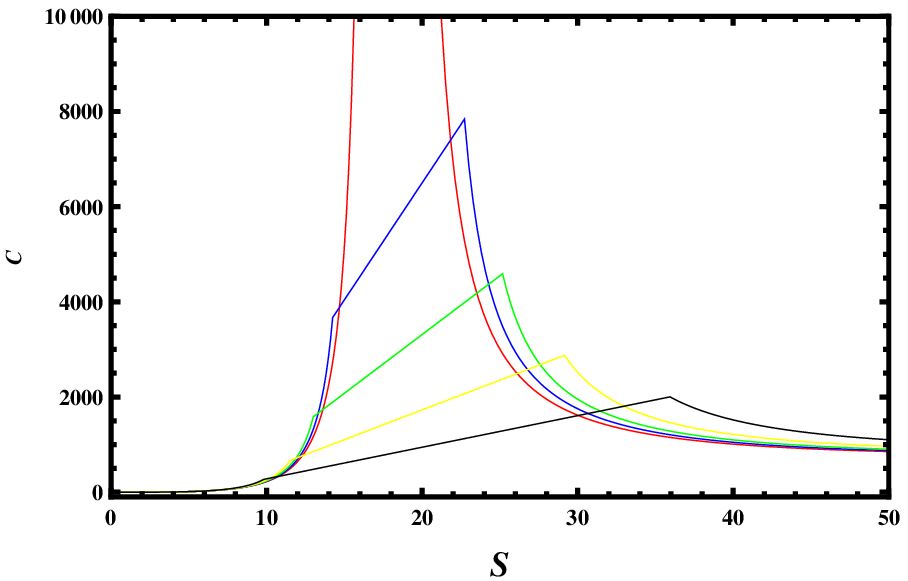}}
\subfigure[]{\label{pCapacityPHs}
\includegraphics[width=6cm]{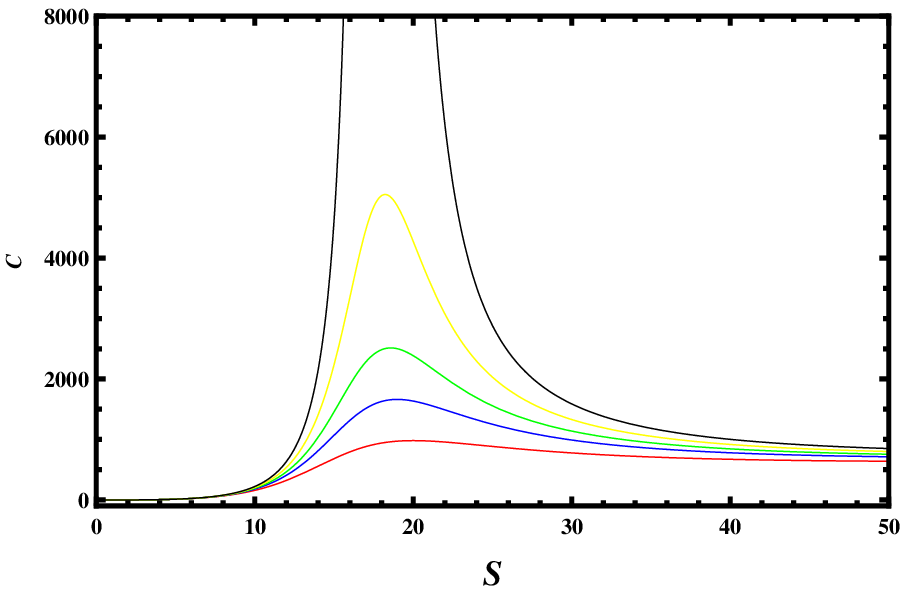}}}
\caption{Heat capacity $C$ vs. entropy $S$ with the charge $Q$=5 when the phase transition is considered. (a) $N^{2}$=3, 3.05, 3.1, 3.2, and 3.4 from top to bottom. (b) $N^{2}$=2.5, 2.7, 2.8, 2.9, and 3 from bottom to top.}\label{ppCapacityPHL}
\end{figure}

\begin{figure}
\center{\subfigure[]{\label{pCapacityPHTL}
\includegraphics[width=6cm]{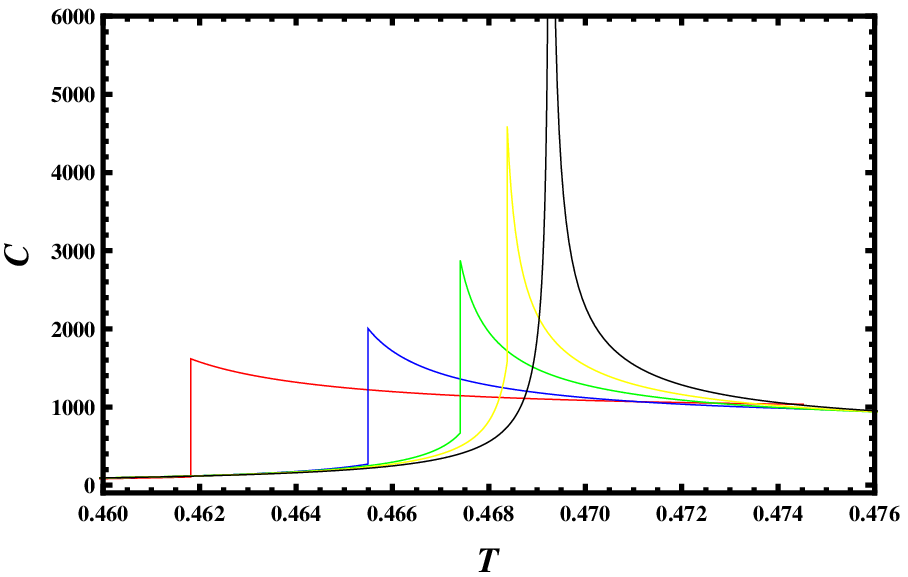}}
\subfigure[]{\label{pCapacityPHTS}
\includegraphics[width=6cm]{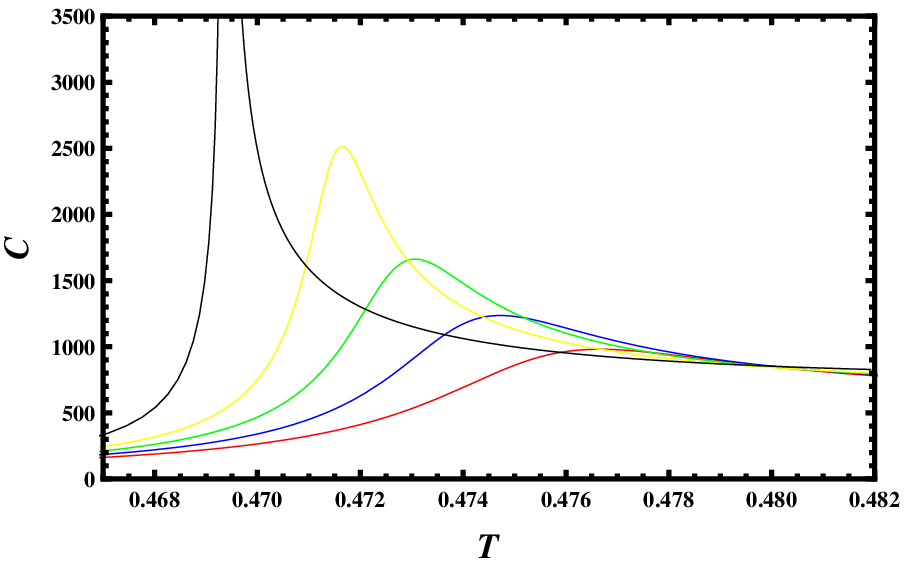}}}
\caption{Heat capacity $C$ vs. temperature $T$ with the charge $Q$=5 when the phase transition is considered. (a) $N^{2}$=3.8, 3.4, 3.2, 3.1, and 3.0 from left to right. (b) $N^{2}$=2.5, 2.6, 2.7, 2.8, and 3.0 from right to left.}\label{ppCapacityPHTL}
\end{figure}

The heat capacity with fixed charge $Q$ and number of colors $N^{2}$ is given by
\begin{eqnarray}
 C=T\bigg(\frac{\partial S}{\partial T}\bigg)_{Q,N^{2}}
   =\frac{3S(6S^{2}+3\pi^{2}N^{4/3}S^{4/3}-\sqrt{2}\pi N Q^{2})}
     {6S^{2}-3\pi^{2}N^{4/3}S^{4/3}+5\sqrt{2}\pi N Q^{2}}.
\end{eqnarray}
At first glance, we can see that this heat capacity vanishes at
\begin{eqnarray}
 Q^{2}=\frac{3(2S^{2}+\pi^{2}N^{4/3}S^{4/3})}{\sqrt{2}\pi N},
\end{eqnarray}
and diverges at
\begin{eqnarray}
 Q^{2}=\frac{3(2S^{2}-\pi^{2}N^{4/3}S^{4/3})}{5\sqrt{2}\pi N}.
\end{eqnarray}
The general view is that the phase transition is linked to the divergent behavior of the heat capacity, so we only focus on its divergent point.

The heat capacity is shown in Fig.~\ref{pCapacitysa} as a function of the entropy $S$ with the charge $Q$=5. For large color number $N^{2}$=3.5 described by the green solid line, we clearly see that there are two divergent points at $S$=12.44 and 28.95. These two points divide the parameter space into three regions. The first and the third regions, i.e., the small and large black hole regions, have positive heat capacity, while the second one denoting the intermediate black hole region has negative heat capacity. As we know, negative values of the heat capacity imply that the thermodynamics is unstable. Thus one may wonder whether this region needs to be removed by some thermodynamic mechanisms. Decreasing the color number, the region of negative heat capacity shrinks. When the color number takes the value of $N^{2}\approx$3, the negative region disappears. The only one divergent point occurs at $S$=17.9. When the entropy $S$ approaches this value from left or right, both values of the heat capacity go to positive infinity. Further decreasing $N^{2}$, we find that the divergent behavior of the heat capacity completely disappears. However, there will be a peak near $S$=17.9.

More interestingly, if we plot the heat capacity as a function of the temperature, there will be some subtle structures emerge, see Fig.~\ref{pCapacityT}. Let us first consider the large color number case with $N^{2}$=3.5 described by the green solid line. In the figure, we use the arrows to indicate the increase of the entropy $S$. For small entropy, the black hole has low temperature. Then the heat capacity and temperature increase with the entropy. The heat capacity goes to positive infinity at $T_{1}$=0.4673. As the entropy increases further, the heat capacity suddenly changes its sign, and increases from negative infinity and then decreases. The temperature also decreases with the entropy. Soon after, it encounters another divergent point, and goes back to the negative infinity at $T_{2}$=0.4631. Next, it changes its sign for the second time and decreases from infinity to some positive values as the entropy increases. The black hole temperature also increases with the entropy. Since the temperature $T_{1}>T_{2}$, the AB line and EF line always have an intersection point. We propose that such behavior implies a first-order phase transition. When we decrease the color number such that $N^{2}\approx$3, the CD line with negative heat capacity will disappear, and the divergent temperatures $T_{1}=T_{2}$. Such situation is quiet similar to that of the vdW fluid system crossing its critical point. Thus we know that it corresponds to a second-order phase transition. When the color number $N^{2}<3$, there is no divergent behavior of the heat capacity. However, there is a peak, and this peak is shifted to high temperature when the number $N^{2}$ decreases.

In the previous section, we have shown that the phase transition point can be determined by the swallow tail behavior of the free energy or the equal area law of the oscillatory curve. So an important question naturally emerges: how the heat capacity behaves when the phase transition is included in? For this purpose, we present the behavior of the heat capacity in Figs.~\ref{ppCapacityPHL} and \ref{ppCapacityPHTL}. The main effect of the phase transition is that we exclude the intermediate black hole region, which is thermodynamically unstable, as well as the super low color number large black hole and super high color number small black hole regions, which are also metastable. In Fig.~\ref{ppCapacityPHL}, we show the heat capacity as a function of the entropy $S$. Figure \ref{pCapacityPHL} is for large $N^{2}$, and Fig.~\ref{pCapacityPHs} for small $N^{2}$. We can see that for $N^{2}>3$, the heat capacity is not smooth for each curve, and in fact it is related to the first-order phase transition. When the color number approaches the critical values, the heat capacity will blow up at the critical entropy, where a second-order phase transition takes place. For small $N^{2}<3$ showed in Fig.~\ref{pCapacityPHs}, the heat capacity has only a peak, and it gets lower and lower with the decrease of the color number. We also plot the heat capacity as a function of the temperature $T$ in Fig.~\ref{ppCapacityPHTL}. From it, we observe the similar situation. For large color number $N^{2}$, the heat capacity does not smoothly change, which implies a first-order phase transition. Note that the phase transition occurs at the same temperature. Similarly, the heat capacity blows up at the critical point. For low color number $N^{2}$, the peak of the heat capacity gets lower and lower, and shifts to higher temperature with the decrease of $N^{2}$.

It is proper to summarize now. The behaviors of the heat capacity described with the green solid lines and blue dashed lines in Fig.~\ref{ppCapacity} indicate the first and second-order phase transitions. So the information of the phase transition indeed encodes in the heat capacity. However, we can not obtain the values for these parameters at the phase transition point only through the heat capacity. While we can read the second-order phase transition information from the divergent behavior of the heat capacity, or
\begin{eqnarray}
 \left(\frac{1}{C}\right)_{Q,N^{2}}=\left(\frac{\partial C}{\partial S}\right)_{Q,N^{2}}=0.
\end{eqnarray}
Nevertheless, thermodynamic instability of the system can be revealed from the sign of the heat capacity.

\subsection{Thermodynamic geometry}

\begin{figure}
\center{\subfigure[$N^{2}$=2.5]{\label{pScalarTa}
\includegraphics[width=6cm]{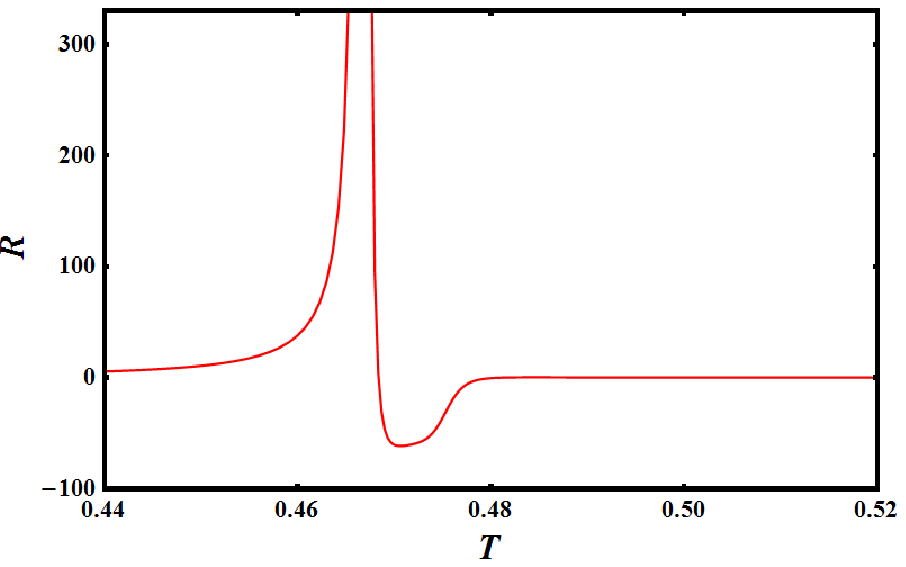}}
\subfigure[$N^{2}$=3]{\label{pScalarTb}
\includegraphics[width=6cm]{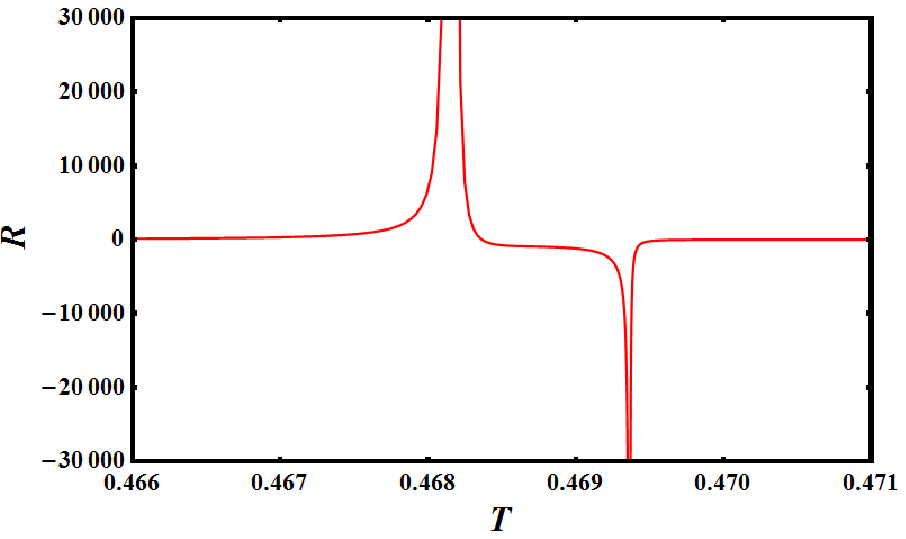}}
\subfigure[$N^{2}$=3.09]{\label{pScalarTc}
\includegraphics[width=6cm]{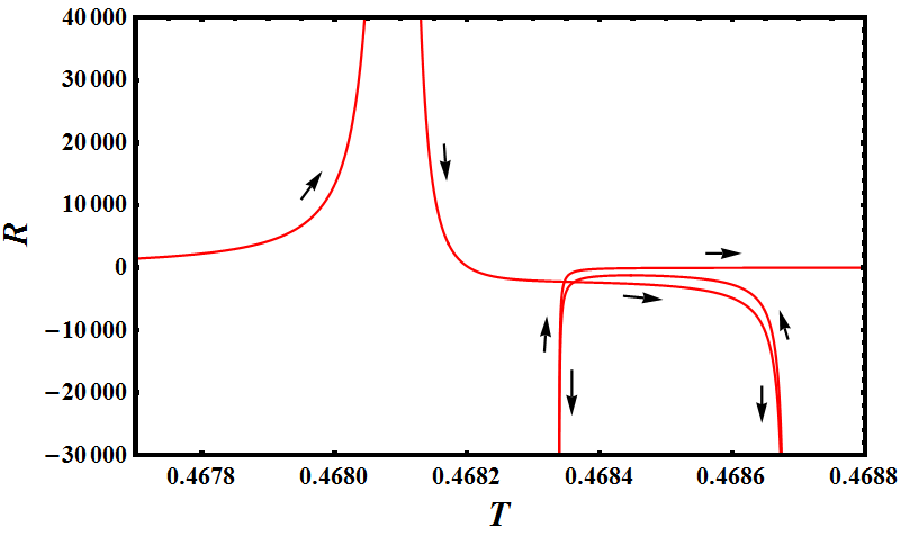}}
\subfigure[$N^{2}$=3.15]{\label{pScalarTd}
\includegraphics[width=6cm]{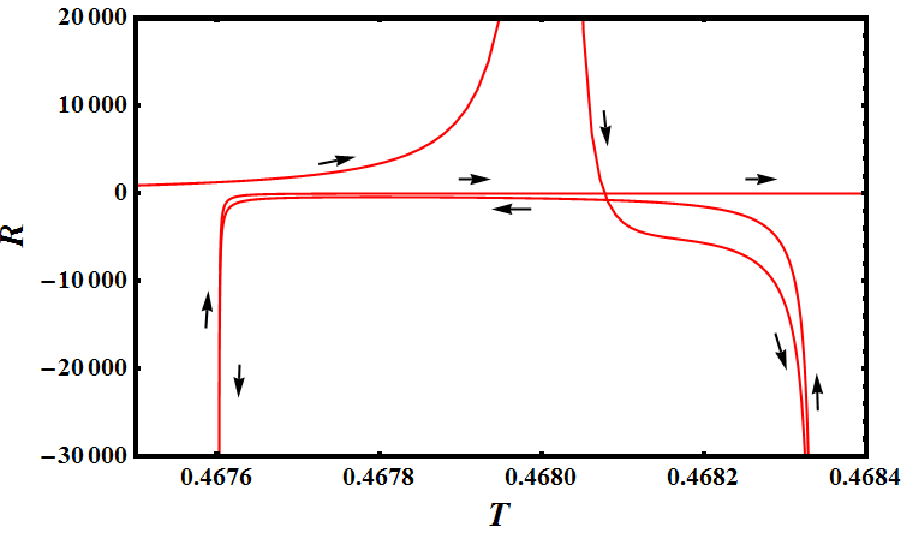}}}
\caption{Curvature scalar $R$ as a function of the temperature $T$. (a) $N^{2}$=2.5, (b) $N^{2}$=3, (c) $N^{2}$=3.09, (d) $N^{2}$=3.15. The arrows indicate the increase of the entropy $S$.}\label{ppScalarTd}
\end{figure}

\begin{figure}
\center{\subfigure[]{\label{pExtremala}
\includegraphics[width=6cm,height=4cm]{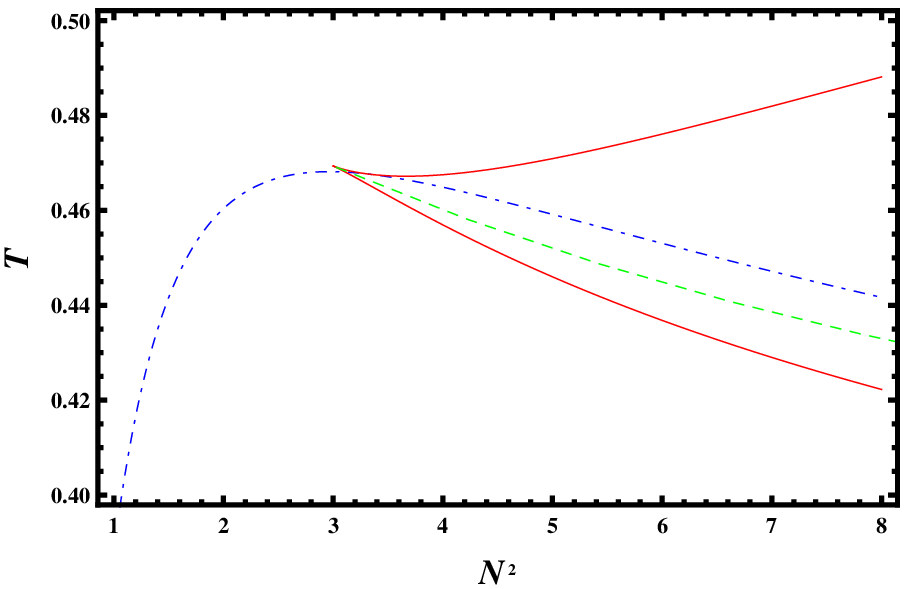}}
\subfigure[]{\label{pExtremalb}
\includegraphics[width=6cm,height=4cm]{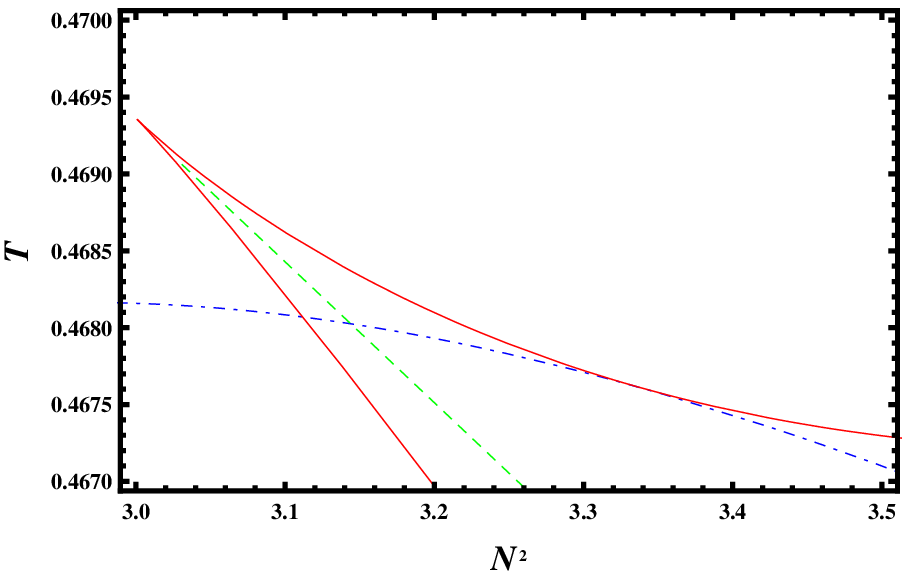}}}
\caption{(a) Divergent behavior of the thermodynamic quantities. The heat capacity diverges at the red solid line. The curvature scalar diverges both at the doted dashed blue line and red solid line. And the phase transition occurs at the dashed green line. The doted dashed blue line and the dashed green line intersect at point (3.112, 0.4681). Black dot denotes the critical point. (b) Magnified image of (a).}\label{pExtremal}
\end{figure}

\begin{figure}
\center{\subfigure[$N^{2}$=3.1]{\label{pScalarPTTa}
\includegraphics[width=6cm]{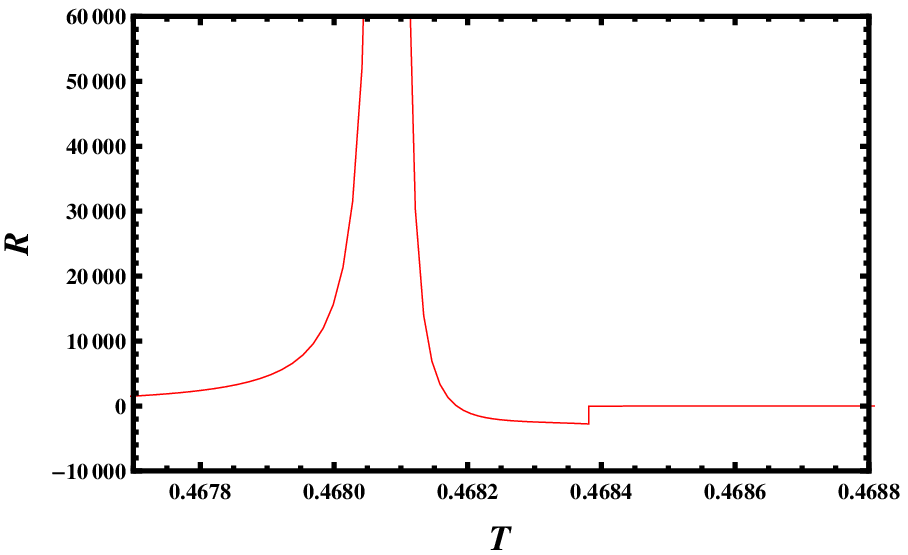}}
\subfigure[$N^{2}$=3.129]{\label{pScalarPTTb}
\includegraphics[width=6cm]{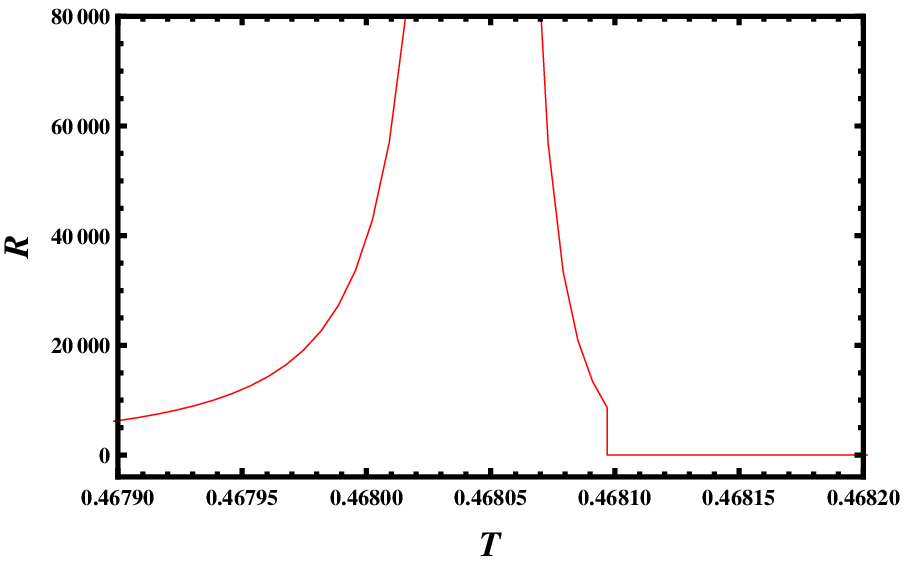}}
\subfigure[$N^{2}$=3.5]{\label{pScalarPTTc}
\includegraphics[width=6cm]{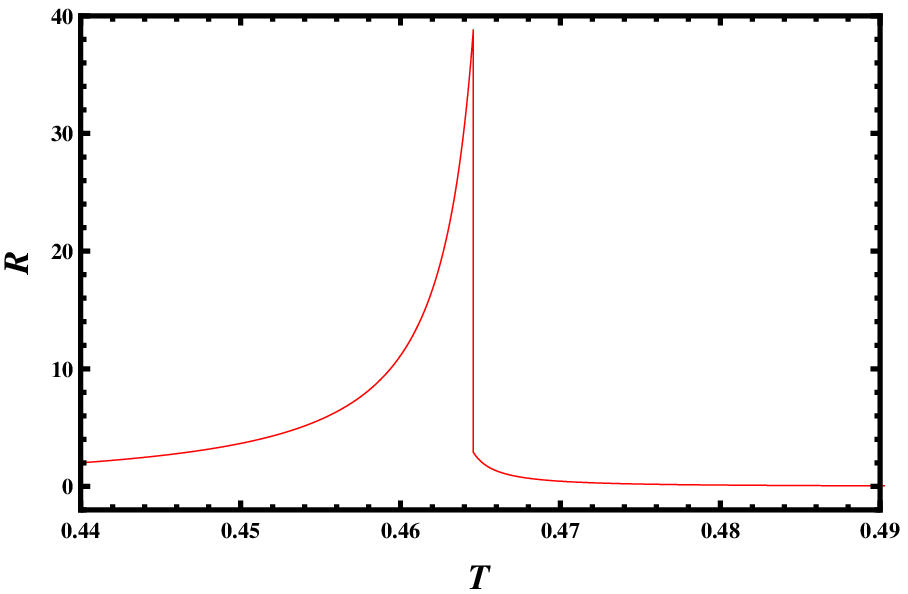}}
\subfigure[$N^{2}$=4]{\label{pScalarPTTd}
\includegraphics[width=6cm]{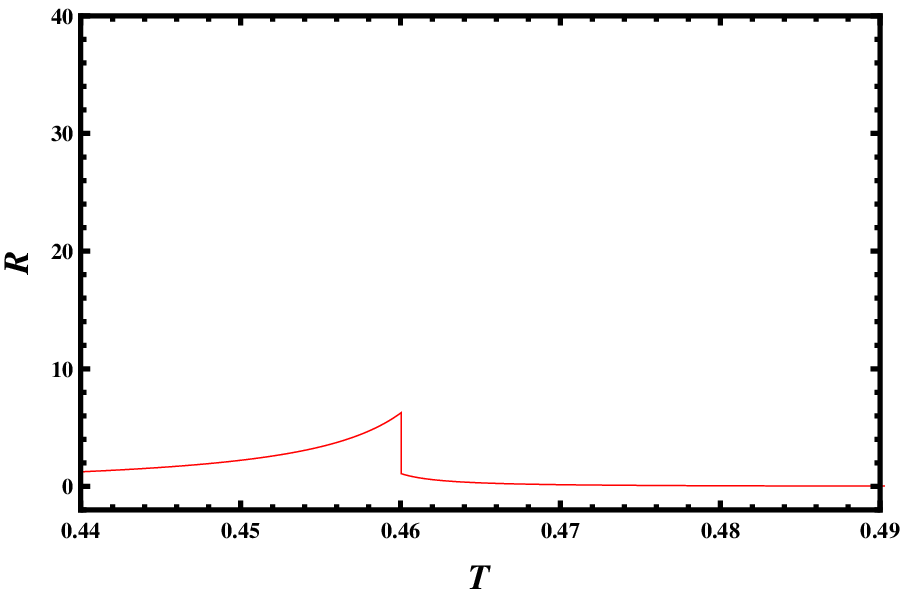}}}
\caption{Scalar $R$ as a function of temperature $T$ when the phase transition is considered. (a) $N^{2}$=3.1, (b) $N^{2}$=3.129, (c) $N^{2}$=3.5, (d) $N^{2}$=4.}\label{ppScalarPTTa}
\end{figure}

One of the aims of the thermodynamic geometry is to investigate the phase transition from the view of the fluctuation theory. Now there are several different thermodynamic geometries, such as the Weinhold geometry \cite{Weinhold}, Ruppeiner geometry \cite{Ruppeiner,Ruppeiner2}, Quevedo geometry \cite{Quevedo}, and so on. Several years ago, the relationship between the curvature scalars constructed from different geometries and different heat capacities were studied in Refs. \cite{Lu,Mansoori,Mirza}. Actually, the difference of these geometries lies in the choice of the thermodynamic potential. However, they are conformally related with each other.

For a thermodynamic system, the phase transition is obtained by the property that it prefers a lower free energy. Therefore, it is natural to chose the free energy $\mathcal{F}(T, N^{2})$ as its thermodynamic potential. For simplicity, we fix the black hole charge $Q$ in our approach. Using the differential form of the free energy (\ref{dfree}), the corresponding metric can be expressed as
\begin{eqnarray}
 ds^{2}=-dSdT+d\mu dN^{2}.
\end{eqnarray}
It is convenient to change the natural variables $(T, N^{2})$ to $(S, N^{2})$. Then the above metric will be of the following form
\begin{eqnarray}
 ds^{2}&=&g_{SS}dS^{2}+2g_{SN^{2}}dSdN^{2}+g_{N^{2}N^{2}}d(N^{2})^{2},\\
 g_{SN^{2}}&=&g_{N^{2}S}=0,\\
 g_{SS}&=&\frac{3\pi^2 N^{4/3} S^{4/3}-5 \sqrt{2}\pi N Q^2-6 S^2}
      {18\sqrt[8]{2}\pi^{7/4} N^{11/12} S^{8/3}},\\
 g_{N^{2}N^{2}}&=&\frac{-285\pi^2 N^{4/3} S^{4/3}-23\sqrt{2}\pi NQ^2+1155 S^2}
    {2304 \sqrt[8]{2}\pi^{7/4} N^{59/12} S^{2/3}}.
\end{eqnarray}
It is clear that this metric is diagonalized. After a simple calculation, we get the curvature scalar
\begin{eqnarray}
 R&=&\Big(-1424\pi^3 N^{7/3} Q^4+46662\sqrt{2}\pi^2 N^{4/3}Q^2 S^2-3915\pi^3 N^{5/3}S^{10/3}-2760\sqrt{2}\pi^4 Q^2
   (N^2 S)^{4/3}\nonumber\\
   &&+16962\pi N Q^4 S^{2/3}+14850\pi\sqrt[3]{N} S^4
   -145926 \sqrt{2}Q^2 S^{8/3}\Big)\times\frac{972\sqrt[8]{2}\pi^{11/4} N^{23/12} S^2}{\chi_{1}^{2}\chi_{2}^{2}},
\end{eqnarray}
where the parameters are given by
\begin{eqnarray}
 \chi_{1}&=&23\sqrt{2}\pi N Q^2-1155 S^2+285\pi^2N^{4/3}S^{4/3},\\
 \chi_{2}&=&5\sqrt{2}\pi N Q^{2}+6S^{2}-3\pi^{2}N^{4/3}S^{4/3}.
\end{eqnarray}
From this expression, we can see that the curvature scalar $R$ diverges at $\chi_{1}=0$ and $\chi_{2}=0$. Moreover, $\chi_{2}$ is just the denominator of the heat capacity, which implies that the divergent point of the heat capacity is also that of the curvature scalar. In order to clearly show the divergent behavior of $R$, we can write it in a more explicit form
\begin{eqnarray}
 R\sim\frac{C^{2}}{\chi_{1}^{2}}.
\end{eqnarray}
We show the behavior of the curvature scalar $R$ as a function of the temperature $T$ in Fig.~\ref{ppScalarTd} for different fixed values of the color number. For small $N^{2}$=2.5, there are one positive divergent point at $T$=0.4668 and one well at $T$=0.4706. When the color number is increased to $N^{2}$=3, the positive divergent point is shifted to $T$=0.4682, and the lowest point of the well goes to negative infinity at $T$=0.4694, which in fact corresponds to the critical point. Increasing the color number slightly to $N^{2}$=3.09, we will get two negative divergent points at $T$=0.4683 and 0.4687, and both of them are larger than the positive divergent point located at $T$=0.4681. However, when the color number gets more larger, such as $N^{2}$=3.15, the positive divergent points will locate between these two negative divergent points, and the corresponding temperatures are $T$=0.4676, 0.4680, 0.4683.
It is also worthwhile noting that the negative divergent points correspond to the divergent points of the heat capacity, i.e., $\chi_{2}=0$, while the positive divergent points correspond to $\chi_{1}=0$.

It is interesting to plot the divergent points of the heat capacity and the curvature scalar in Fig.~\ref{pExtremal} in the $T$-$N^{2}$ chart. The divergent points of the heat capacity are denoted with the red solid lines. These two lines intersect at the critical point $(N^{2}, T)=(3.0011, 0.4694)$. For the color number $N^{2}<3.0011$, the heat capacity has no divergent behavior. On the other hand, the curvature scalar also diverges at the red solid lines, as well as the blue dot dashed line. The green dashed line denotes the first-order phase transition curve. The magnified image near the critical point is given in Fig.~\ref{pExtremalb}. From it, we can find the subtle divergent behaviors of the heat capacity and the curvature scalar near the critical point.

Now, we are interested in the behavior of the curvature scalar when the phase transition is considered. The result is showed in Fig.~\ref{ppScalarPTTa} for the color number $N^{2}>N_{c}^{2}$. From it, we see that the negative infinity is replaced with a finite change. However, the positive infinity exists for $N^{2}$=3.1 and 3.129, see Figs.~\ref{pScalarPTTa} and \ref{pScalarPTTb}. While it disappears in Fig.~\ref{pScalarPTTc} and Fig.~\ref{pScalarPTTd}. A careful check implies that the critical point for the vanished positive infinity occurs at $N^{2}$=3.112.

In summary, using the divergent behavior of the curvature scalar, we can find some information of the black hole phase transition. In particular, similar to the heat capacity, the critical point of the phase transition can be obtained by solving
\begin{eqnarray}
 \left(\frac{1}{R}\right)_{Q,N^{2}}
    =\left(\frac{\partial R}{\partial T}\right)_{Q,N^{2}}=0.
\end{eqnarray}
It is also important to note that, there is another divergent behavior of the curvature scalar, i.e., $\chi_{1}$=0. This result suggests that the divergent behavior of the curvature scalar is useful on revealing the phase transition information, but sometimes it is problematic.

\subsection{Along the coexistence curve}

In the $N^{2}$-$T$ phase diagram, we see that with the increase of the temperature, the color number monotonically decreases until the critical point is approached. And here we are interested in how the heat capacity and curvature scalar behave when the parameters vary along the coexistence curve of the saturated small and large black hole curves?

\begin{figure}
\center{\subfigure[$N^{2}$=5]{\label{pAlongHCa}
\includegraphics[width=6cm]{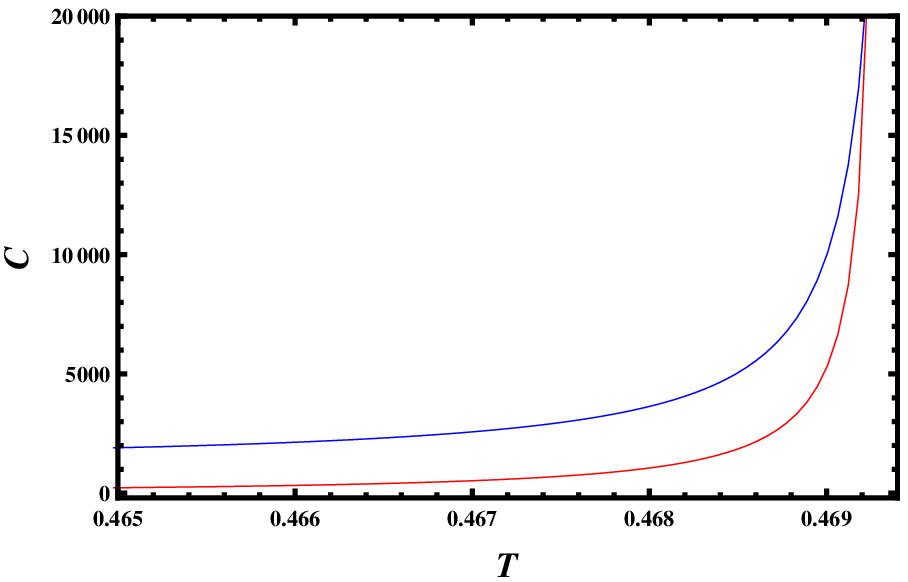}}
\subfigure[$N^{2}$=10]{\label{pAlongHCb}
\includegraphics[width=6cm]{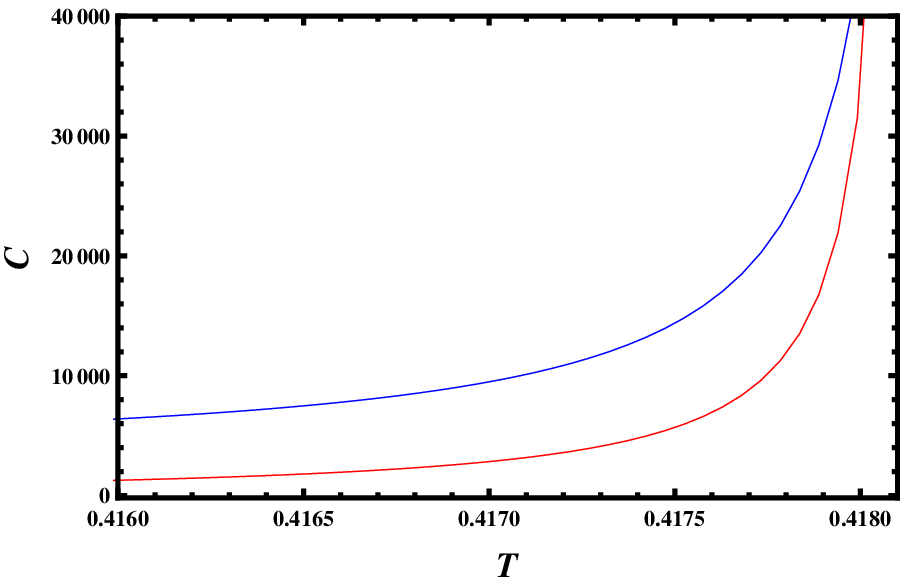}}}
\caption{Heat capacity along the coexistence curve. (a) $N^{2}$=5. (b) $N^{2}$=10.}\label{ppAlongHCb}
\end{figure}

\begin{figure}
\center{\subfigure[$N^{2}$=5]{\label{pAlongGeomea}
\includegraphics[width=6cm]{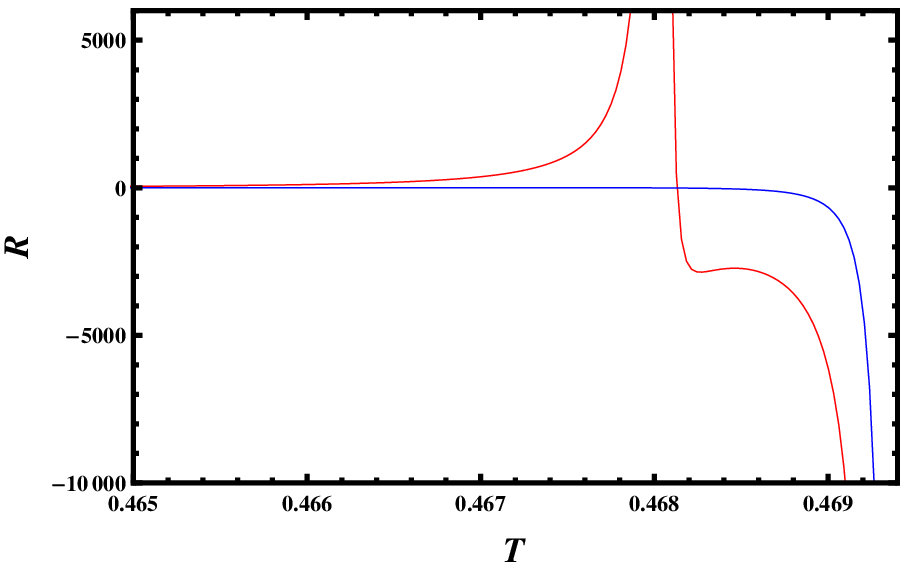}}
\subfigure[$N^{2}$=10]{\label{pAlongGeomeb}
\includegraphics[width=6cm]{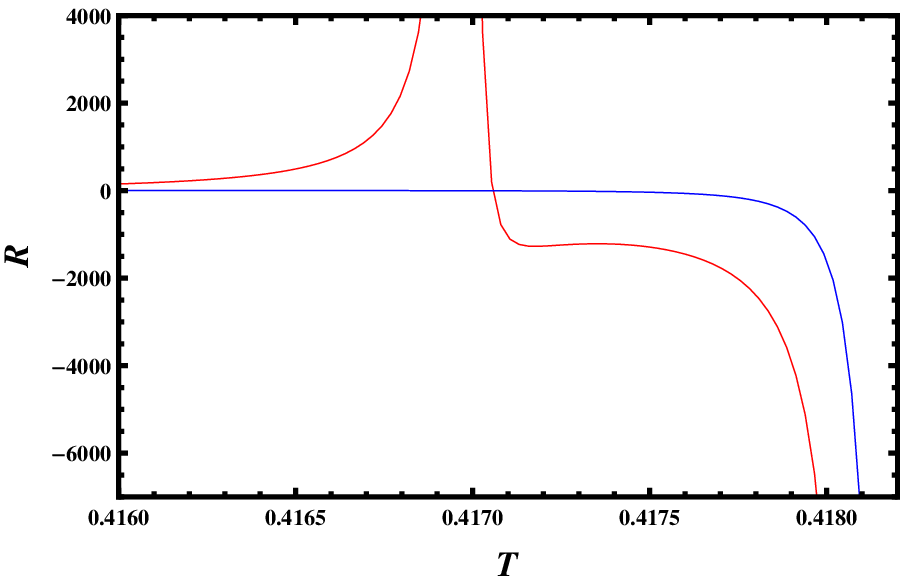}}}
\caption{Curvature scalar $R$ along the coexistence curve. (a) $N^{2}$=5. (b) $N^{2}$=10.}\label{ppAlongGeomeb}
\end{figure}

Since there is no the analytic expression of the coexistence curve, we only numerically study it. The heat capacity is given in Fig.~\ref{ppAlongHCb}. The case for the saturated large black hole is described by the top blue line, and that for the saturated small black hole is described by the bottom red line for the color number $N^{2}$=5 and 10, respectively. From the figure, we can see that the saturated large black holes always have high heat capacity than the small ones. More importantly, both of them go to positive infinity when the critical point is approached. This is also a significant feature of the critical point. The behavior of the curvature scalar along the coexistence is depicted in Fig.~\ref{ppAlongGeomeb} for the color number $N^{2}$=5 and 10, respectively. We can see that the curvature scalar of the saturated large black hole first slowly varies with the temperature, and then rapidly decreases to negative infinity near the critical point. On the other hand, the curvature scalar of the saturated small black hole first increases to positive infinity, then decreases to a finite negative value, and finally goes to negative infinity when the critical temperature is arrived. Thus the curvature scalar along the saturated small black hole curve is problematic.

Another interesting issue is to numerically calculate the critical exponents. For the heat capacity and curvature  scalar, we expect, near the critical point,
\begin{eqnarray}
 C&\sim&3^{5/6}(2/5\pi)^{1/3}Q^{\frac{4}{3}}\times (1-\tilde{T})^{-\alpha},\\
 R&\sim&162\times6^{5/6}5^{-7/12}\pi^{2/3}Q^{-\frac{7}{6}}\times (1-\tilde{T})^{-\beta},
\end{eqnarray}
or, in the form
\begin{eqnarray}
 \log |C|&\sim&-\alpha \log|1-\tilde{T}|+\texttt{constant},\\
 \log |R|&\sim&-\beta \log |1-\tilde{T}|+\texttt{constant}.
\end{eqnarray}
Then fitting the data when the system approaches the critical point along the saturated small and large black hole curves, we have
\begin{equation}
\log|C|=\begin{cases}
 -1.0045  \log |1-\tilde{T}| -0.5780,&  $for saturated large black hole$,\\
 -1.0042  \log |1-\tilde{T}| -0.8787,&  $for saturated small black hole$.\\
\end{cases}
\end{equation}
and
\begin{equation}
\log\mid R\mid=\begin{cases}
 -2.1268  \log\mid 1-\tilde{T}\mid-12.8589,&  $for saturated large black hole$,\\
 -2.1280  \log\mid 1-\tilde{T}\mid-12.8762,&  $for saturated small black hole$.\\
\end{cases}
\end{equation}
Clearly, we approximately have $\alpha$=1, which is just the same as the one from the mean field theory. Moreover, for the curvature scalar, its exponent is of about $\beta$=2, which is the same as that of the correlation length. So this result may guide us on constructing the correlation length \cite{Ruppeiner2}, and some properties of the micro structure of a black hole may be revealed through it.

A brief summary is that both the heat capacity and the curvature scalar go to infinity at the critical point and share the same critical exponents as the mean field theory, which, in some sense, can be treated as the features of the black hole critical phenomena.

\section{Conclusions and discussions}
\label{Conclusion}

In this paper, by treating the cosmological constant as the color number $N^2$ in the boundary gauge theory and its conjugate quantity as the associated chemical potential $\mu$, we have studied the thermodynamics and critical phenomena for the charged AdS black hole.

The result shows that there exists a small-large black hole phase transition, which is similar to the vdW fluid. Especially, with the help of the $N^{2}$-$\mu$ criticality, the phase transition was investigated. We clearly showed the equal area law in the $N^{2}$-$\mu$ chart and the swallow tail behavior of the Helmholtz free energy. And by utilizing them, we got the phase diagram and coexistence curve for the phase transition, see Fig.~\ref{pnmuptaa}. These results indicate that the phase transition prefers low temperature $T$ and large color number $N^{2}$. Moreover, we also calculated the latent heat for the phase transition. With the increase of the temperature $T$, it firstly starts at a very large value and then rapidly decreases. At the critical point, the latent heat approaches zero. Such image confirms that the small-large black hole phase transition is a first-order phase transition, while the critical point is a second-order one.

In the reduced parameter space, all the reduced thermodynamic quantities were found to be charge-independent. And the phase transition was examined. In particular, different black hole phases were displayed. For example, in the $\tilde{N}^{2}$-$\tilde{\mu}$ chart (see Fig.~\ref{pReducedNmuphase}), six black hole phases participate in.

On the other hand, two important quantities, the heat capacity and thermodynamic curvature scalar, were studied aiming at revealing the black hole phase transition. At first, we plotted the heat capacity as a function of the entropy with fixed charge. For large color number $N^{2}$, the heat capacity has two divergent points and a parameter region of negative value indicating thermodynamic instability. When decreasing $N^{2}$, the negative range narrows and disappears for the critical case. Further decreasing $N^{2}$, the divergent point disappears, while with a peak left. When the small-large black hole phase transition is considered, there is only a divergent point, and which is nothing but the critical point. One interesting thermodynamic phenomenon was revealed when we showed the heat capacity as a function of the temperature, see Fig.~\ref{pCapacityT}. There appears a two-turn-back behavior at the divergent points for the large color number, which we suggest as a remarkable phenomenon for the existence of the first-order black hole phase transition. It is also worthwhile noting that such behavior disappears when the first-order black hole phase transition is included in. So, we can conclude that the first-order black hole phase transition can be revealed from the behavior of the heat capacity. However, the values of these quantities cannot be determined only by the heat capacity. Nevertheless, the critical point linked to a second-order phase transition can be determined by the heat capacity.

With the help of the free energy, we constructed a thermodynamic geometry. The curvature scalar was obtained. From the curvature scalar, the information of the phase transition can also be revealed as that of the heat capacity. This originates from that one of the divergent points of the curvature scalar is the same as that of the heat capacity. Moreover, there exists another divergent point for the curvature scalar. However, it is not related to the phase transition.

Interestingly, we also studied the behaviors of the heat capacity and curvature scalar along the coexistence curve. The heat capacity goes to positive infinity both for the saturated small and large black holes when the critical temperature is approached. While the curvature scalar goes to negative infinity as the critical temperature is reached. However, the curvature scalar along the saturated small black hole curve is problematic for there is a divergent point, but it has nothing to do with the phase transition. Near the critical temperature, we found that the critical exponents for the heat capacity and the curvature scalar are approximately $\alpha=1$ and $\beta=2$. These results have an important application on revealing the micro structure of the black holes. And according to the AdS/CFT correspondence, our results may provide a preliminary understanding on the cosmological constant in the dual field theory.

\section*{Acknowledgements}
This work was supported by the National Natural Science Foundation of China (Grants No. 11675064, No. 11522541, No. 11205074 and No. 11375075), and the Fundamental Research Funds for the Central Universities (No. lzujbky-2016-121 and lzujbky-2016-k04).

\end{document}